\documentclass[longauth]{aaV2}

\usepackage{euclid}
\usepackage{graphicx}
\usepackage{natbib}
\usepackage{scalerel}
\usepackage{dcolumn}
\usepackage[nolist,nohyperlinks]{acronym}
\usepackage{lastpage}

\usepackage[table]{xcolor}

\usepackage{txfonts}
\usepackage[pdfencoding=auto,psdextra]{hyperref}
\hypersetup{
    colorlinks=true,
    linkcolor=blue,
    filecolor=magenta,      
    urlcolor=blue,
    citecolor=blue
}
\makeatletter
\renewcommand*\aa@pageof{, page \thepage{} of \pageref*{LastPage}}
\makeatother

\usepackage[utf8]{inputenc}

\usepackage[mathlines,switch, modulo]{lineno}

\usepackage{orcidlink} 
\newcommand{\orcid}[1]{\orcidlink{#1}}

\begin{document}


\acrodef{NISP}{Near-Infrared Spectrometer and Photometer}
\acrodef{LED}{light-emitting diode}
\acrodef{NIR}{near-infrared}
\acrodef{FPA}{focal plane array}


%
%
\title{\Euclid. IV. The NISP Calibration Unit} 



\author{Euclid Collaboration: F.~Hormuth\inst{\ref{aff1},\ref{aff2}}
\and K.~Jahnke\orcid{0000-0003-3804-2137}\thanks{\email{jahnke@mpia.de}}\inst{\ref{aff2}}
\and M.~Schirmer\orcid{0000-0003-2568-9994}\inst{\ref{aff2}}
\and C.~G.-Y.~Lee\inst{\ref{aff3},\ref{aff4}}
\and T.~Scott\inst{\ref{aff3}}
\and R.~Barbier\inst{\ref{aff5}}
\and S.~Ferriol\inst{\ref{aff5}}
\and W.~Gillard\orcid{0000-0003-4744-9748}\inst{\ref{aff6}}
\and F.~Grupp\inst{\ref{aff7},\ref{aff8}}
\and R.~Holmes\inst{\ref{aff9}}
\and W.~Holmes\inst{\ref{aff10}}
\and B.~Kubik\orcid{0009-0006-5823-4880}\inst{\ref{aff5}}
\and J.~Macias-Perez\orcid{0000-0002-5385-2763}\inst{\ref{aff11}}
\and M.~Laurent\inst{\ref{aff12}}
\and J.~Marpaud\inst{\ref{aff11}}
\and M.~Marton\inst{\ref{aff11}}
\and E.~Medinaceli\orcid{0000-0002-4040-7783}\inst{\ref{aff13}}
\and G.~Morgante\inst{\ref{aff13}}
\and R.~Toledo-Moreo\orcid{0000-0002-2997-4859}\inst{\ref{aff14}}
\and M.~Trifoglio\orcid{0000-0002-2505-3630}\inst{\ref{aff13}}
\and Hans-Walter~Rix\orcid{0000-0003-4996-9069}\inst{\ref{aff2}}
\and A.~Secroun\orcid{0000-0003-0505-3710}\inst{\ref{aff6}}
\and M.~Seiffert\orcid{0000-0002-7536-9393}\inst{\ref{aff10}}
\and P.~Stassi\orcid{0000-0001-5584-8410}\inst{\ref{aff11}}
\and S.~Wachter\inst{\ref{aff15}}
\and C.~M.~Gutierrez\inst{\ref{aff16}}
\and C.~Vescovi\inst{\ref{aff11}}
\and A.~Amara\inst{\ref{aff17}}
\and S.~Andreon\orcid{0000-0002-2041-8784}\inst{\ref{aff18}}
\and N.~Auricchio\orcid{0000-0003-4444-8651}\inst{\ref{aff13}}
\and C.~Baccigalupi\orcid{0000-0002-8211-1630}\inst{\ref{aff19},\ref{aff20},\ref{aff21},\ref{aff22}}
\and M.~Baldi\orcid{0000-0003-4145-1943}\inst{\ref{aff23},\ref{aff13},\ref{aff24}}
\and A.~Balestra\orcid{0000-0002-6967-261X}\inst{\ref{aff25}}
\and S.~Bardelli\orcid{0000-0002-8900-0298}\inst{\ref{aff13}}
\and P.~Battaglia\orcid{0000-0002-7337-5909}\inst{\ref{aff13}}
\and R.~Bender\orcid{0000-0001-7179-0626}\inst{\ref{aff7},\ref{aff8}}
\and C.~Bodendorf\inst{\ref{aff7}}
\and D.~Bonino\orcid{0000-0002-3336-9977}\inst{\ref{aff26}}
\and E.~Branchini\orcid{0000-0002-0808-6908}\inst{\ref{aff27},\ref{aff28},\ref{aff18}}
\and M.~Brescia\orcid{0000-0001-9506-5680}\inst{\ref{aff29},\ref{aff30},\ref{aff31}}
\and J.~Brinchmann\orcid{0000-0003-4359-8797}\inst{\ref{aff32}}
\and S.~Camera\orcid{0000-0003-3399-3574}\inst{\ref{aff33},\ref{aff34},\ref{aff26}}
\and V.~Capobianco\orcid{0000-0002-3309-7692}\inst{\ref{aff26}}
\and C.~Carbone\orcid{0000-0003-0125-3563}\inst{\ref{aff35}}
\and V.~F.~Cardone\inst{\ref{aff36},\ref{aff37}}
\and J.~Carretero\orcid{0000-0002-3130-0204}\inst{\ref{aff38},\ref{aff39}}
\and R.~Casas\orcid{0000-0002-8165-5601}\inst{\ref{aff40},\ref{aff41}}
\and S.~Casas\orcid{0000-0002-4751-5138}\inst{\ref{aff42}}
\and M.~Castellano\orcid{0000-0001-9875-8263}\inst{\ref{aff36}}
\and G.~Castignani\orcid{0000-0001-6831-0687}\inst{\ref{aff13}}
\and S.~Cavuoti\orcid{0000-0002-3787-4196}\inst{\ref{aff30},\ref{aff31}}
\and A.~Cimatti\inst{\ref{aff43}}
\and C.~Colodro-Conde\inst{\ref{aff44}}
\and G.~Congedo\orcid{0000-0003-2508-0046}\inst{\ref{aff45}}
\and C.~J.~Conselice\orcid{0000-0003-1949-7638}\inst{\ref{aff46}}
\and L.~Conversi\orcid{0000-0002-6710-8476}\inst{\ref{aff47},\ref{aff48}}
\and Y.~Copin\orcid{0000-0002-5317-7518}\inst{\ref{aff5}}
\and L.~Corcione\orcid{0000-0002-6497-5881}\inst{\ref{aff26}}
\and F.~Courbin\orcid{0000-0003-0758-6510}\inst{\ref{aff49}}
\and H.~M.~Courtois\orcid{0000-0003-0509-1776}\inst{\ref{aff50}}
\and A.~Da~Silva\orcid{0000-0002-6385-1609}\inst{\ref{aff51},\ref{aff52}}
\and H.~Degaudenzi\orcid{0000-0002-5887-6799}\inst{\ref{aff53}}
\and G.~De~Lucia\orcid{0000-0002-6220-9104}\inst{\ref{aff20}}
\and J.~Dinis\inst{\ref{aff51},\ref{aff52}}
\and M.~Douspis\orcid{0000-0003-4203-3954}\inst{\ref{aff54}}
\and F.~Dubath\orcid{0000-0002-6533-2810}\inst{\ref{aff53}}
\and F.~Ducret\inst{\ref{aff12}}
\and X.~Dupac\inst{\ref{aff48}}
\and S.~Dusini\orcid{0000-0002-1128-0664}\inst{\ref{aff55}}
\and M.~Fabricius\orcid{0000-0002-7025-6058}\inst{\ref{aff7},\ref{aff8}}
\and M.~Farina\orcid{0000-0002-3089-7846}\inst{\ref{aff56}}
\and S.~Farrens\orcid{0000-0002-9594-9387}\inst{\ref{aff57}}
\and F.~Faustini\orcid{0000-0001-6274-5145}\inst{\ref{aff58},\ref{aff36}}
\and S.~Fotopoulou\orcid{0000-0002-9686-254X}\inst{\ref{aff59}}
\and N.~Fourmanoit\inst{\ref{aff6}}
\and M.~Frailis\orcid{0000-0002-7400-2135}\inst{\ref{aff20}}
\and E.~Franceschi\orcid{0000-0002-0585-6591}\inst{\ref{aff13}}
\and P.~Franzetti\inst{\ref{aff35}}
\and M.~Fumana\orcid{0000-0001-6787-5950}\inst{\ref{aff35}}
\and S.~Galeotta\orcid{0000-0002-3748-5115}\inst{\ref{aff20}}
\and B.~Garilli\orcid{0000-0001-7455-8750}\inst{\ref{aff35}}
\and K.~George\orcid{0000-0002-1734-8455}\inst{\ref{aff8}}
\and B.~Gillis\orcid{0000-0002-4478-1270}\inst{\ref{aff45}}
\and C.~Giocoli\orcid{0000-0002-9590-7961}\inst{\ref{aff13},\ref{aff60}}
\and A.~Grazian\orcid{0000-0002-5688-0663}\inst{\ref{aff25}}
\and L.~Guzzo\orcid{0000-0001-8264-5192}\inst{\ref{aff61},\ref{aff18}}
\and S.~V.~H.~Haugan\orcid{0000-0001-9648-7260}\inst{\ref{aff62}}
\and H.~Hoekstra\orcid{0000-0002-0641-3231}\inst{\ref{aff63}}
\and I.~Hook\orcid{0000-0002-2960-978X}\inst{\ref{aff64}}
\and A.~Hornstrup\orcid{0000-0002-3363-0936}\inst{\ref{aff65},\ref{aff66}}
\and P.~Hudelot\inst{\ref{aff67}}
\and M.~Jhabvala\inst{\ref{aff68}}
\and E.~Keih\"anen\orcid{0000-0003-1804-7715}\inst{\ref{aff69}}
\and S.~Kermiche\orcid{0000-0002-0302-5735}\inst{\ref{aff6}}
\and A.~Kiessling\orcid{0000-0002-2590-1273}\inst{\ref{aff10}}
\and M.~Kilbinger\orcid{0000-0001-9513-7138}\inst{\ref{aff57}}
\and T.~Kitching\orcid{0000-0002-4061-4598}\inst{\ref{aff70}}
\and R.~Kohley\inst{\ref{aff48}}
\and M.~K\"ummel\orcid{0000-0003-2791-2117}\inst{\ref{aff8}}
\and M.~Kunz\orcid{0000-0002-3052-7394}\inst{\ref{aff71}}
\and H.~Kurki-Suonio\orcid{0000-0002-4618-3063}\inst{\ref{aff72},\ref{aff73}}
\and D.~Le~Mignant\orcid{0000-0002-5339-5515}\inst{\ref{aff12}}
\and S.~Ligori\orcid{0000-0003-4172-4606}\inst{\ref{aff26}}
\and P.~B.~Lilje\orcid{0000-0003-4324-7794}\inst{\ref{aff62}}
\and V.~Lindholm\orcid{0000-0003-2317-5471}\inst{\ref{aff72},\ref{aff73}}
\and I.~Lloro\inst{\ref{aff74}}
\and G.~Mainetti\inst{\ref{aff75}}
\and E.~Maiorano\orcid{0000-0003-2593-4355}\inst{\ref{aff13}}
\and O.~Mansutti\orcid{0000-0001-5758-4658}\inst{\ref{aff20}}
\and S.~Marcin\inst{\ref{aff76}}
\and O.~Marggraf\orcid{0000-0001-7242-3852}\inst{\ref{aff77}}
\and K.~Markovic\orcid{0000-0001-6764-073X}\inst{\ref{aff10}}
\and M.~Martinelli\orcid{0000-0002-6943-7732}\inst{\ref{aff36},\ref{aff37}}
\and N.~Martinet\orcid{0000-0003-2786-7790}\inst{\ref{aff12}}
\and F.~Marulli\orcid{0000-0002-8850-0303}\inst{\ref{aff78},\ref{aff13},\ref{aff24}}
\and R.~Massey\orcid{0000-0002-6085-3780}\inst{\ref{aff79}}
\and S.~Maurogordato\inst{\ref{aff80}}
\and H.~J.~McCracken\orcid{0000-0002-9489-7765}\inst{\ref{aff67}}
\and S.~Mei\orcid{0000-0002-2849-559X}\inst{\ref{aff81}}
\and M.~Melchior\inst{\ref{aff76}}
\and Y.~Mellier\inst{\ref{aff82},\ref{aff67}}
\and M.~Meneghetti\orcid{0000-0003-1225-7084}\inst{\ref{aff13},\ref{aff24}}
\and E.~Merlin\orcid{0000-0001-6870-8900}\inst{\ref{aff36}}
\and G.~Meylan\inst{\ref{aff49}}
\and J.~J.~Mohr\orcid{0000-0002-6875-2087}\inst{\ref{aff8},\ref{aff7}}
\and M.~Moresco\orcid{0000-0002-7616-7136}\inst{\ref{aff78},\ref{aff13}}
\and P.~W.~Morris\orcid{0000-0002-5186-4381}\inst{\ref{aff83}}
\and L.~Moscardini\orcid{0000-0002-3473-6716}\inst{\ref{aff78},\ref{aff13},\ref{aff24}}
\and E.~Munari\orcid{0000-0002-1751-5946}\inst{\ref{aff20},\ref{aff22}}
\and R.~Nakajima\inst{\ref{aff77}}
\and C.~Neissner\orcid{0000-0001-8524-4968}\inst{\ref{aff84},\ref{aff39}}
\and R.~C.~Nichol\inst{\ref{aff17}}
\and S.-M.~Niemi\inst{\ref{aff85}}
\and J.~W.~Nightingale\orcid{0000-0002-8987-7401}\inst{\ref{aff86},\ref{aff87}}
\and C.~Padilla\orcid{0000-0001-7951-0166}\inst{\ref{aff84}}
\and K.~Paech\orcid{0000-0003-0625-2367}\inst{\ref{aff7}}
\and S.~Paltani\orcid{0000-0002-8108-9179}\inst{\ref{aff53}}
\and F.~Pasian\orcid{0000-0002-4869-3227}\inst{\ref{aff20}}
\and K.~Pedersen\inst{\ref{aff88}}
\and W.~J.~Percival\orcid{0000-0002-0644-5727}\inst{\ref{aff89},\ref{aff90},\ref{aff91}}
\and V.~Pettorino\inst{\ref{aff85}}
\and S.~Pires\orcid{0000-0002-0249-2104}\inst{\ref{aff57}}
\and G.~Polenta\orcid{0000-0003-4067-9196}\inst{\ref{aff58}}
\and M.~Poncet\inst{\ref{aff92}}
\and L.~A.~Popa\inst{\ref{aff93}}
\and F.~Raison\orcid{0000-0002-7819-6918}\inst{\ref{aff7}}
\and R.~Rebolo\inst{\ref{aff44},\ref{aff94}}
\and A.~Renzi\orcid{0000-0001-9856-1970}\inst{\ref{aff95},\ref{aff55}}
\and J.~Rhodes\inst{\ref{aff10}}
\and G.~Riccio\inst{\ref{aff30}}
\and E.~Romelli\orcid{0000-0003-3069-9222}\inst{\ref{aff20}}
\and M.~Roncarelli\orcid{0000-0001-9587-7822}\inst{\ref{aff13}}
\and E.~Rossetti\inst{\ref{aff23}}
\and B.~Rusholme\orcid{0000-0001-7648-4142}\inst{\ref{aff96}}
\and R.~Saglia\orcid{0000-0003-0378-7032}\inst{\ref{aff8},\ref{aff7}}
\and Z.~Sakr\orcid{0000-0002-4823-3757}\inst{\ref{aff97},\ref{aff98},\ref{aff99}}
\and A.~G.~S\'anchez\orcid{0000-0003-1198-831X}\inst{\ref{aff7}}
\and D.~Sapone\orcid{0000-0001-7089-4503}\inst{\ref{aff100}}
\and B.~Sartoris\orcid{0000-0003-1337-5269}\inst{\ref{aff8},\ref{aff20}}
\and M.~Sauvage\orcid{0000-0002-0809-2574}\inst{\ref{aff57}}
\and J.~A.~Schewtschenko\inst{\ref{aff45}}
\and P.~Schneider\orcid{0000-0001-8561-2679}\inst{\ref{aff77}}
\and T.~Schrabback\orcid{0000-0002-6987-7834}\inst{\ref{aff101}}
\and E.~Sefusatti\orcid{0000-0003-0473-1567}\inst{\ref{aff20},\ref{aff22},\ref{aff21}}
\and G.~Seidel\orcid{0000-0003-2907-353X}\inst{\ref{aff2}}
\and S.~Serrano\orcid{0000-0002-0211-2861}\inst{\ref{aff40},\ref{aff41},\ref{aff102}}
\and C.~Sirignano\orcid{0000-0002-0995-7146}\inst{\ref{aff95},\ref{aff55}}
\and G.~Sirri\orcid{0000-0003-2626-2853}\inst{\ref{aff24}}
\and G.~Smadja\inst{\ref{aff5}}
\and L.~Stanco\orcid{0000-0002-9706-5104}\inst{\ref{aff55}}
\and J.~Steinwagner\inst{\ref{aff7}}
\and P.~Tallada-Cresp\'{i}\orcid{0000-0002-1336-8328}\inst{\ref{aff38},\ref{aff39}}
\and D.~Tavagnacco\orcid{0000-0001-7475-9894}\inst{\ref{aff20}}
\and A.~N.~Taylor\inst{\ref{aff45}}
\and H.~I.~Teplitz\orcid{0000-0002-7064-5424}\inst{\ref{aff103}}
\and I.~Tereno\inst{\ref{aff51},\ref{aff104}}
\and F.~Torradeflot\orcid{0000-0003-1160-1517}\inst{\ref{aff39},\ref{aff38}}
\and I.~Tutusaus\orcid{0000-0002-3199-0399}\inst{\ref{aff98}}
\and L.~Valenziano\orcid{0000-0002-1170-0104}\inst{\ref{aff13},\ref{aff105}}
\and T.~Vassallo\orcid{0000-0001-6512-6358}\inst{\ref{aff8},\ref{aff20}}
\and A.~Veropalumbo\orcid{0000-0003-2387-1194}\inst{\ref{aff18},\ref{aff28}}
\and Y.~Wang\orcid{0000-0002-4749-2984}\inst{\ref{aff103}}
\and J.~Weller\orcid{0000-0002-8282-2010}\inst{\ref{aff8},\ref{aff7}}
\and A.~Zacchei\orcid{0000-0003-0396-1192}\inst{\ref{aff20},\ref{aff22}}
\and G.~Zamorani\orcid{0000-0002-2318-301X}\inst{\ref{aff13}}
\and F.~M.~Zerbi\inst{\ref{aff18}}
\and E.~Zucca\orcid{0000-0002-5845-8132}\inst{\ref{aff13}}
\and A.~Biviano\orcid{0000-0002-0857-0732}\inst{\ref{aff20},\ref{aff22}}
\and M.~Bolzonella\orcid{0000-0003-3278-4607}\inst{\ref{aff13}}
\and A.~Boucaud\orcid{0000-0001-7387-2633}\inst{\ref{aff81}}
\and E.~Bozzo\orcid{0000-0002-8201-1525}\inst{\ref{aff53}}
\and C.~Burigana\orcid{0000-0002-3005-5796}\inst{\ref{aff106},\ref{aff105}}
\and M.~Calabrese\orcid{0000-0002-2637-2422}\inst{\ref{aff107},\ref{aff35}}
\and D.~Di~Ferdinando\inst{\ref{aff24}}
\and J.~A.~Escartin~Vigo\inst{\ref{aff7}}
\and R.~Farinelli\inst{\ref{aff13}}
\and J.~Gracia-Carpio\inst{\ref{aff7}}
\and M.~V.~Kazandjian\inst{\ref{aff63},\ref{aff108}}
\and N.~Mauri\orcid{0000-0001-8196-1548}\inst{\ref{aff43},\ref{aff24}}
\and V.~Scottez\inst{\ref{aff82},\ref{aff109}}
\and M.~Tenti\orcid{0000-0002-4254-5901}\inst{\ref{aff24}}
\and M.~Viel\orcid{0000-0002-2642-5707}\inst{\ref{aff22},\ref{aff20},\ref{aff19},\ref{aff21},\ref{aff110}}
\and M.~Wiesmann\orcid{0009-0000-8199-5860}\inst{\ref{aff62}}
\and Y.~Akrami\orcid{0000-0002-2407-7956}\inst{\ref{aff111},\ref{aff112}}
\and V.~Allevato\orcid{0000-0001-7232-5152}\inst{\ref{aff30}}
\and S.~Anselmi\orcid{0000-0002-3579-9583}\inst{\ref{aff55},\ref{aff95},\ref{aff113}}
\and E.~Aubourg\orcid{0000-0002-5592-023X}\inst{\ref{aff81},\ref{aff114}}
\and M.~Ballardini\orcid{0000-0003-4481-3559}\inst{\ref{aff115},\ref{aff13},\ref{aff116}}
\and M.~Bethermin\orcid{0000-0002-3915-2015}\inst{\ref{aff117},\ref{aff12}}
\and A.~Blanchard\orcid{0000-0001-8555-9003}\inst{\ref{aff98}}
\and L.~Blot\orcid{0000-0002-9622-7167}\inst{\ref{aff118},\ref{aff113}}
\and S.~Borgani\orcid{0000-0001-6151-6439}\inst{\ref{aff119},\ref{aff22},\ref{aff20},\ref{aff21}}
\and A.~S.~Borlaff\orcid{0000-0003-3249-4431}\inst{\ref{aff120},\ref{aff121}}
\and E.~Borsato\inst{\ref{aff95},\ref{aff55}}
\and S.~Bruton\orcid{0000-0002-6503-5218}\inst{\ref{aff122}}
\and R.~Cabanac\orcid{0000-0001-6679-2600}\inst{\ref{aff98}}
\and A.~Calabro\orcid{0000-0003-2536-1614}\inst{\ref{aff36}}
\and G.~Canas-Herrera\orcid{0000-0003-2796-2149}\inst{\ref{aff85},\ref{aff123}}
\and A.~Cappi\inst{\ref{aff13},\ref{aff80}}
\and C.~S.~Carvalho\inst{\ref{aff104}}
\and P.~Casenove\inst{\ref{aff92}}
\and T.~Castro\orcid{0000-0002-6292-3228}\inst{\ref{aff20},\ref{aff21},\ref{aff22},\ref{aff110}}
\and K.~C.~Chambers\orcid{0000-0001-6965-7789}\inst{\ref{aff124}}
\and Y.~Charles\inst{\ref{aff12}}
\and S.~Contarini\orcid{0000-0002-9843-723X}\inst{\ref{aff7},\ref{aff78}}
\and A.~R.~Cooray\orcid{0000-0002-3892-0190}\inst{\ref{aff125}}
\and O.~Cucciati\orcid{0000-0002-9336-7551}\inst{\ref{aff13}}
\and S.~Davini\orcid{0000-0003-3269-1718}\inst{\ref{aff28}}
\and B.~De~Caro\inst{\ref{aff55},\ref{aff95}}
\and S.~de~la~Torre\inst{\ref{aff12}}
\and G.~Desprez\inst{\ref{aff126}}
\and A.~D\'iaz-S\'anchez\orcid{0000-0003-0748-4768}\inst{\ref{aff127}}
\and J.~J.~Diaz\inst{\ref{aff16}}
\and S.~Di~Domizio\orcid{0000-0003-2863-5895}\inst{\ref{aff27},\ref{aff28}}
\and H.~Dole\orcid{0000-0002-9767-3839}\inst{\ref{aff54}}
\and S.~Escoffier\orcid{0000-0002-2847-7498}\inst{\ref{aff6}}
\and A.~G.~Ferrari\orcid{0009-0005-5266-4110}\inst{\ref{aff43},\ref{aff24}}
\and P.~G.~Ferreira\orcid{0000-0002-3021-2851}\inst{\ref{aff128}}
\and I.~Ferrero\orcid{0000-0002-1295-1132}\inst{\ref{aff62}}
\and F.~Finelli\orcid{0000-0002-6694-3269}\inst{\ref{aff13},\ref{aff105}}
\and A.~Fontana\orcid{0000-0003-3820-2823}\inst{\ref{aff36}}
\and F.~Fornari\orcid{0000-0003-2979-6738}\inst{\ref{aff105}}
\and L.~Gabarra\orcid{0000-0002-8486-8856}\inst{\ref{aff128}}
\and K.~Ganga\orcid{0000-0001-8159-8208}\inst{\ref{aff81}}
\and J.~Garc\'ia-Bellido\orcid{0000-0002-9370-8360}\inst{\ref{aff111}}
\and E.~Gaztanaga\orcid{0000-0001-9632-0815}\inst{\ref{aff41},\ref{aff40},\ref{aff129}}
\and F.~Giacomini\orcid{0000-0002-3129-2814}\inst{\ref{aff24}}
\and G.~Gozaliasl\orcid{0000-0002-0236-919X}\inst{\ref{aff130},\ref{aff72}}
\and A.~Hall\orcid{0000-0002-3139-8651}\inst{\ref{aff45}}
\and W.~G.~Hartley\inst{\ref{aff53}}
\and H.~Hildebrandt\orcid{0000-0002-9814-3338}\inst{\ref{aff131}}
\and J.~Hjorth\orcid{0000-0002-4571-2306}\inst{\ref{aff132}}
\and M.~Huertas-Company\orcid{0000-0002-1416-8483}\inst{\ref{aff44},\ref{aff16},\ref{aff133},\ref{aff134}}
\and O.~Ilbert\orcid{0000-0002-7303-4397}\inst{\ref{aff12}}
\and J.~Jacobson\inst{\ref{aff96}}
\and A.~Jimenez~Mu\~noz\orcid{0009-0004-5252-185X}\inst{\ref{aff11}}
\and S.~Joudaki\orcid{0000-0001-8820-673X}\inst{\ref{aff129}}
\and J.~J.~E.~Kajava\orcid{0000-0002-3010-8333}\inst{\ref{aff135},\ref{aff136}}
\and V.~Kansal\orcid{0000-0002-4008-6078}\inst{\ref{aff137},\ref{aff138}}
\and D.~Karagiannis\orcid{0000-0002-4927-0816}\inst{\ref{aff139},\ref{aff140}}
\and C.~C.~Kirkpatrick\inst{\ref{aff69}}
\and F.~Laudisio\inst{\ref{aff55}}
\and L.~Legrand\orcid{0000-0003-0610-5252}\inst{\ref{aff141}}
\and G.~Libet\inst{\ref{aff92}}
\and A.~Loureiro\orcid{0000-0002-4371-0876}\inst{\ref{aff142},\ref{aff143}}
\and G.~Maggio\orcid{0000-0003-4020-4836}\inst{\ref{aff20}}
\and M.~Magliocchetti\orcid{0000-0001-9158-4838}\inst{\ref{aff56}}
\and C.~Mancini\orcid{0000-0002-4297-0561}\inst{\ref{aff35}}
\and F.~Mannucci\orcid{0000-0002-4803-2381}\inst{\ref{aff144}}
\and R.~Maoli\orcid{0000-0002-6065-3025}\inst{\ref{aff145},\ref{aff36}}
\and C.~J.~A.~P.~Martins\orcid{0000-0002-4886-9261}\inst{\ref{aff146},\ref{aff32}}
\and S.~Matthew\inst{\ref{aff45}}
\and L.~Maurin\orcid{0000-0002-8406-0857}\inst{\ref{aff54}}
\and R.~B.~Metcalf\orcid{0000-0003-3167-2574}\inst{\ref{aff78},\ref{aff13}}
\and M.~Miluzio\inst{\ref{aff48},\ref{aff147}}
\and C.~Moretti\orcid{0000-0003-3314-8936}\inst{\ref{aff19},\ref{aff110},\ref{aff20},\ref{aff22},\ref{aff21}}
\and S.~Nadathur\orcid{0000-0001-9070-3102}\inst{\ref{aff129}}
\and Nicholas~A.~Walton\orcid{0000-0003-3983-8778}\inst{\ref{aff148}}
\and L.~Patrizii\inst{\ref{aff24}}
\and A.~Pezzotta\orcid{0000-0003-0726-2268}\inst{\ref{aff7}}
\and M.~P\"ontinen\orcid{0000-0001-5442-2530}\inst{\ref{aff72}}
\and V.~Popa\inst{\ref{aff93}}
\and C.~Porciani\orcid{0000-0002-7797-2508}\inst{\ref{aff77}}
\and D.~Potter\orcid{0000-0002-0757-5195}\inst{\ref{aff149}}
\and I.~Risso\orcid{0000-0003-2525-7761}\inst{\ref{aff150}}
\and P.-F.~Rocci\inst{\ref{aff54}}
\and R.~P.~Rollins\orcid{0000-0003-1291-1023}\inst{\ref{aff45}}
\and M.~Sahl\'en\orcid{0000-0003-0973-4804}\inst{\ref{aff151}}
\and C.~Scarlata\orcid{0000-0002-9136-8876}\inst{\ref{aff122}}
\and A.~Schneider\orcid{0000-0001-7055-8104}\inst{\ref{aff149}}
\and M.~Schultheis\inst{\ref{aff80}}
\and M.~Sereno\orcid{0000-0003-0302-0325}\inst{\ref{aff13},\ref{aff24}}
\and A.~Shulevski\orcid{0000-0002-1827-0469}\inst{\ref{aff152},\ref{aff153},\ref{aff154}}
\and A.~Silvestri\orcid{0000-0001-6904-5061}\inst{\ref{aff123}}
\and P.~Simon\inst{\ref{aff77}}
\and A.~Spurio~Mancini\orcid{0000-0001-5698-0990}\inst{\ref{aff155},\ref{aff70}}
\and J.~Stadel\orcid{0000-0001-7565-8622}\inst{\ref{aff149}}
\and C.~Tao\orcid{0000-0001-7961-8177}\inst{\ref{aff6}}
\and G.~Testera\inst{\ref{aff28}}
\and R.~Teyssier\orcid{0000-0001-7689-0933}\inst{\ref{aff156}}
\and S.~Toft\orcid{0000-0003-3631-7176}\inst{\ref{aff66},\ref{aff157}}
\and S.~Tosi\orcid{0000-0002-7275-9193}\inst{\ref{aff27},\ref{aff28},\ref{aff18}}
\and A.~Troja\orcid{0000-0003-0239-4595}\inst{\ref{aff95},\ref{aff55}}
\and M.~Tucci\inst{\ref{aff53}}
\and C.~Valieri\inst{\ref{aff24}}
\and J.~Valiviita\orcid{0000-0001-6225-3693}\inst{\ref{aff72},\ref{aff73}}
\and D.~Vergani\orcid{0000-0003-0898-2216}\inst{\ref{aff13}}
\and G.~Verza\orcid{0000-0002-1886-8348}\inst{\ref{aff158},\ref{aff159}}
\and L.~Zalesky\orcid{0000-0001-5680-2326}\inst{\ref{aff124}}
\and M.~Archidiacono\orcid{0000-0003-4952-9012}\inst{\ref{aff61},\ref{aff160}}
\and F.~Atrio-Barandela\orcid{0000-0002-2130-2513}\inst{\ref{aff161}}
\and T.~Bouvard\orcid{0009-0002-7959-312X}\inst{\ref{aff162}}
\and F.~Caro\inst{\ref{aff36}}
\and P.~Dimauro\orcid{0000-0001-7399-2854}\inst{\ref{aff36},\ref{aff163}}
\and Y.~Fang\inst{\ref{aff8}}
\and A.~M.~N.~Ferguson\inst{\ref{aff45}}
\and A.~Finoguenov\orcid{0000-0002-4606-5403}\inst{\ref{aff72}}
\and T.~Gasparetto\orcid{0000-0002-7913-4866}\inst{\ref{aff20}}
\and A.~M.~C.~Le~Brun\orcid{0000-0002-0936-4594}\inst{\ref{aff113}}
\and J.~Le~Graet\orcid{0000-0001-6523-7971}\inst{\ref{aff6}}
\and T.~I.~Liaudat\orcid{0000-0002-9104-314X}\inst{\ref{aff114}}
\and A.~Montoro\orcid{0000-0003-4730-8590}\inst{\ref{aff41},\ref{aff40}}
\and C.~Murray\inst{\ref{aff81}}
\and M.~Oguri\orcid{0000-0003-3484-399X}\inst{\ref{aff164},\ref{aff165}}
\and L.~Pagano\orcid{0000-0003-1820-5998}\inst{\ref{aff115},\ref{aff116}}
\and D.~Paoletti\orcid{0000-0003-4761-6147}\inst{\ref{aff13},\ref{aff105}}
\and E.~Sarpa\orcid{0000-0002-1256-655X}\inst{\ref{aff19},\ref{aff110},\ref{aff21}}
\and K.~Tanidis\inst{\ref{aff128}}
\and F.~Vernizzi\orcid{0000-0003-3426-2802}\inst{\ref{aff166}}
\and A.~Viitanen\orcid{0000-0001-9383-786X}\inst{\ref{aff69},\ref{aff36}}
\and I.~Kova{\v{c}}i{\'{c}}\orcid{0000-0001-6751-3263}\inst{\ref{aff167}}
\and J.~Lesgourgues\orcid{0000-0001-7627-353X}\inst{\ref{aff42}}
\and J.~Mart\'{i}n-Fleitas\orcid{0000-0002-8594-569X}\inst{\ref{aff168}}
\and A.~Mora\orcid{0000-0002-1922-8529}\inst{\ref{aff168}}}
										   
\institute{Felix Hormuth Engineering, Goethestr. 17, 69181 Leimen, Germany\label{aff1}
\and
Max-Planck-Institut f\"ur Astronomie, K\"onigstuhl 17, 69117 Heidelberg, Germany\label{aff2}
\and
von Hoerner \& Sulger GmbH, Schlossplatz 8, 68723 Schwetzingen, Germany\label{aff3}
\and
European Organisation for the Exploitation of Meteorological Satellites (EUMETSAT), Eumetsat Allee 1, D-64295 Darmstadt, Germany\label{aff4}
\and
Universit\'e Claude Bernard Lyon 1, CNRS/IN2P3, IP2I Lyon, UMR 5822, Villeurbanne, F-69100, France\label{aff5}
\and
Aix-Marseille Universit\'e, CNRS/IN2P3, CPPM, Marseille, France\label{aff6}
\and
Max Planck Institute for Extraterrestrial Physics, Giessenbachstr. 1, 85748 Garching, Germany\label{aff7}
\and
Universit\"ats-Sternwarte M\"unchen, Fakult\"at f\"ur Physik, Ludwig-Maximilians-Universit\"at M\"unchen, Scheinerstrasse 1, 81679 M\"unchen, Germany\label{aff8}
\and
Surrey Satellite Technology Limited, Tycho House, 20 Stephenson Road, Surrey Research Park, Guildford, GU2 7YE, UK\label{aff9}
\and
Jet Propulsion Laboratory, California Institute of Technology, 4800 Oak Grove Drive, Pasadena, CA, 91109, USA\label{aff10}
\and
Univ. Grenoble Alpes, CNRS, Grenoble INP, LPSC-IN2P3, 53, Avenue des Martyrs, 38000, Grenoble, France\label{aff11}
\and
Aix-Marseille Universit\'e, CNRS, CNES, LAM, Marseille, France\label{aff12}
\and
INAF-Osservatorio di Astrofisica e Scienza dello Spazio di Bologna, Via Piero Gobetti 93/3, 40129 Bologna, Italy\label{aff13}
\and
Universidad Polit\'ecnica de Cartagena, Departamento de Electr\'onica y Tecnolog\'ia de Computadoras,  Plaza del Hospital 1, 30202 Cartagena, Spain\label{aff14}
\and
Carnegie Observatories, Pasadena, CA 91101, USA\label{aff15}
\and
Instituto de Astrof\'isica de Canarias (IAC); Departamento de Astrof\'isica, Universidad de La Laguna (ULL), 38200, La Laguna, Tenerife, Spain\label{aff16}
\and
School of Mathematics and Physics, University of Surrey, Guildford, Surrey, GU2 7XH, UK\label{aff17}
\and
INAF-Osservatorio Astronomico di Brera, Via Brera 28, 20122 Milano, Italy\label{aff18}
\and
SISSA, International School for Advanced Studies, Via Bonomea 265, 34136 Trieste TS, Italy\label{aff19}
\and
INAF-Osservatorio Astronomico di Trieste, Via G. B. Tiepolo 11, 34143 Trieste, Italy\label{aff20}
\and
INFN, Sezione di Trieste, Via Valerio 2, 34127 Trieste TS, Italy\label{aff21}
\and
IFPU, Institute for Fundamental Physics of the Universe, via Beirut 2, 34151 Trieste, Italy\label{aff22}
\and
Dipartimento di Fisica e Astronomia, Universit\`a di Bologna, Via Gobetti 93/2, 40129 Bologna, Italy\label{aff23}
\and
INFN-Sezione di Bologna, Viale Berti Pichat 6/2, 40127 Bologna, Italy\label{aff24}
\and
INAF-Osservatorio Astronomico di Padova, Via dell'Osservatorio 5, 35122 Padova, Italy\label{aff25}
\and
INAF-Osservatorio Astrofisico di Torino, Via Osservatorio 20, 10025 Pino Torinese (TO), Italy\label{aff26}
\and
Dipartimento di Fisica, Universit\`a di Genova, Via Dodecaneso 33, 16146, Genova, Italy\label{aff27}
\and
INFN-Sezione di Genova, Via Dodecaneso 33, 16146, Genova, Italy\label{aff28}
\and
Department of Physics "E. Pancini", University Federico II, Via Cinthia 6, 80126, Napoli, Italy\label{aff29}
\and
INAF-Osservatorio Astronomico di Capodimonte, Via Moiariello 16, 80131 Napoli, Italy\label{aff30}
\and
INFN section of Naples, Via Cinthia 6, 80126, Napoli, Italy\label{aff31}
\and
Instituto de Astrof\'isica e Ci\^encias do Espa\c{c}o, Universidade do Porto, CAUP, Rua das Estrelas, PT4150-762 Porto, Portugal\label{aff32}
\and
Dipartimento di Fisica, Universit\`a degli Studi di Torino, Via P. Giuria 1, 10125 Torino, Italy\label{aff33}
\and
INFN-Sezione di Torino, Via P. Giuria 1, 10125 Torino, Italy\label{aff34}
\and
INAF-IASF Milano, Via Alfonso Corti 12, 20133 Milano, Italy\label{aff35}
\and
INAF-Osservatorio Astronomico di Roma, Via Frascati 33, 00078 Monteporzio Catone, Italy\label{aff36}
\and
INFN-Sezione di Roma, Piazzale Aldo Moro, 2 - c/o Dipartimento di Fisica, Edificio G. Marconi, 00185 Roma, Italy\label{aff37}
\and
Centro de Investigaciones Energ\'eticas, Medioambientales y Tecnol\'ogicas (CIEMAT), Avenida Complutense 40, 28040 Madrid, Spain\label{aff38}
\and
Port d'Informaci\'{o} Cient\'{i}fica, Campus UAB, C. Albareda s/n, 08193 Bellaterra (Barcelona), Spain\label{aff39}
\and
Institut d'Estudis Espacials de Catalunya (IEEC),  Edifici RDIT, Campus UPC, 08860 Castelldefels, Barcelona, Spain\label{aff40}
\and
Institute of Space Sciences (ICE, CSIC), Campus UAB, Carrer de Can Magrans, s/n, 08193 Barcelona, Spain\label{aff41}
\and
Institute for Theoretical Particle Physics and Cosmology (TTK), RWTH Aachen University, 52056 Aachen, Germany\label{aff42}
\and
Dipartimento di Fisica e Astronomia "Augusto Righi" - Alma Mater Studiorum Universit\`a di Bologna, Viale Berti Pichat 6/2, 40127 Bologna, Italy\label{aff43}
\and
Instituto de Astrof\'isica de Canarias, Calle V\'ia L\'actea s/n, 38204, San Crist\'obal de La Laguna, Tenerife, Spain\label{aff44}
\and
Institute for Astronomy, University of Edinburgh, Royal Observatory, Blackford Hill, Edinburgh EH9 3HJ, UK\label{aff45}
\and
Jodrell Bank Centre for Astrophysics, Department of Physics and Astronomy, University of Manchester, Oxford Road, Manchester M13 9PL, UK\label{aff46}
\and
European Space Agency/ESRIN, Largo Galileo Galilei 1, 00044 Frascati, Roma, Italy\label{aff47}
\and
ESAC/ESA, Camino Bajo del Castillo, s/n., Urb. Villafranca del Castillo, 28692 Villanueva de la Ca\~nada, Madrid, Spain\label{aff48}
\and
Institute of Physics, Laboratory of Astrophysics, Ecole Polytechnique F\'ed\'erale de Lausanne (EPFL), Observatoire de Sauverny, 1290 Versoix, Switzerland\label{aff49}
\and
UCB Lyon 1, CNRS/IN2P3, IUF, IP2I Lyon, 4 rue Enrico Fermi, 69622 Villeurbanne, France\label{aff50}
\and
Departamento de F\'isica, Faculdade de Ci\^encias, Universidade de Lisboa, Edif\'icio C8, Campo Grande, PT1749-016 Lisboa, Portugal\label{aff51}
\and
Instituto de Astrof\'isica e Ci\^encias do Espa\c{c}o, Faculdade de Ci\^encias, Universidade de Lisboa, Campo Grande, 1749-016 Lisboa, Portugal\label{aff52}
\and
Department of Astronomy, University of Geneva, ch. d'Ecogia 16, 1290 Versoix, Switzerland\label{aff53}
\and
Universit\'e Paris-Saclay, CNRS, Institut d'astrophysique spatiale, 91405, Orsay, France\label{aff54}
\and
INFN-Padova, Via Marzolo 8, 35131 Padova, Italy\label{aff55}
\and
INAF-Istituto di Astrofisica e Planetologia Spaziali, via del Fosso del Cavaliere, 100, 00100 Roma, Italy\label{aff56}
\and
Universit\'e Paris-Saclay, Universit\'e Paris Cit\'e, CEA, CNRS, AIM, 91191, Gif-sur-Yvette, France\label{aff57}
\and
Space Science Data Center, Italian Space Agency, via del Politecnico snc, 00133 Roma, Italy\label{aff58}
\and
School of Physics, HH Wills Physics Laboratory, University of Bristol, Tyndall Avenue, Bristol, BS8 1TL, UK\label{aff59}
\and
Istituto Nazionale di Fisica Nucleare, Sezione di Bologna, Via Irnerio 46, 40126 Bologna, Italy\label{aff60}
\and
Dipartimento di Fisica "Aldo Pontremoli", Universit\`a degli Studi di Milano, Via Celoria 16, 20133 Milano, Italy\label{aff61}
\and
Institute of Theoretical Astrophysics, University of Oslo, P.O. Box 1029 Blindern, 0315 Oslo, Norway\label{aff62}
\and
Leiden Observatory, Leiden University, Einsteinweg 55, 2333 CC Leiden, The Netherlands\label{aff63}
\and
Department of Physics, Lancaster University, Lancaster, LA1 4YB, UK\label{aff64}
\and
Technical University of Denmark, Elektrovej 327, 2800 Kgs. Lyngby, Denmark\label{aff65}
\and
Cosmic Dawn Center (DAWN), Denmark\label{aff66}
\and
Institut d'Astrophysique de Paris, UMR 7095, CNRS, and Sorbonne Universit\'e, 98 bis boulevard Arago, 75014 Paris, France\label{aff67}
\and
NASA Goddard Space Flight Center, Greenbelt, MD 20771, USA\label{aff68}
\and
Department of Physics and Helsinki Institute of Physics, Gustaf H\"allstr\"omin katu 2, 00014 University of Helsinki, Finland\label{aff69}
\and
Mullard Space Science Laboratory, University College London, Holmbury St Mary, Dorking, Surrey RH5 6NT, UK\label{aff70}
\and
Universit\'e de Gen\`eve, D\'epartement de Physique Th\'eorique and Centre for Astroparticle Physics, 24 quai Ernest-Ansermet, CH-1211 Gen\`eve 4, Switzerland\label{aff71}
\and
Department of Physics, P.O. Box 64, 00014 University of Helsinki, Finland\label{aff72}
\and
Helsinki Institute of Physics, Gustaf H{\"a}llstr{\"o}min katu 2, University of Helsinki, Helsinki, Finland\label{aff73}
\and
NOVA optical infrared instrumentation group at ASTRON, Oude Hoogeveensedijk 4, 7991PD, Dwingeloo, The Netherlands\label{aff74}
\and
Centre de Calcul de l'IN2P3/CNRS, 21 avenue Pierre de Coubertin 69627 Villeurbanne Cedex, France\label{aff75}
\and
University of Applied Sciences and Arts of Northwestern Switzerland, School of Engineering, 5210 Windisch, Switzerland\label{aff76}
\and
Universit\"at Bonn, Argelander-Institut f\"ur Astronomie, Auf dem H\"ugel 71, 53121 Bonn, Germany\label{aff77}
\and
Dipartimento di Fisica e Astronomia "Augusto Righi" - Alma Mater Studiorum Universit\`a di Bologna, via Piero Gobetti 93/2, 40129 Bologna, Italy\label{aff78}
\and
Department of Physics, Centre for Extragalactic Astronomy, Durham University, South Road, DH1 3LE, UK\label{aff79}
\and
Universit\'e C\^{o}te d'Azur, Observatoire de la C\^{o}te d'Azur, CNRS, Laboratoire Lagrange, Bd de l'Observatoire, CS 34229, 06304 Nice cedex 4, France\label{aff80}
\and
Universit\'e Paris Cit\'e, CNRS, Astroparticule et Cosmologie, 75013 Paris, France\label{aff81}
\and
Institut d'Astrophysique de Paris, 98bis Boulevard Arago, 75014, Paris, France\label{aff82}
\and
California institute of Technology, 1200 E California Blvd, Pasadena, CA 91125, USA\label{aff83}
\and
Institut de F\'{i}sica d'Altes Energies (IFAE), The Barcelona Institute of Science and Technology, Campus UAB, 08193 Bellaterra (Barcelona), Spain\label{aff84}
\and
European Space Agency/ESTEC, Keplerlaan 1, 2201 AZ Noordwijk, The Netherlands\label{aff85}
\and
School of Mathematics, Statistics and Physics, Newcastle University, Herschel Building, Newcastle-upon-Tyne, NE1 7RU, UK\label{aff86}
\and
Department of Physics, Institute for Computational Cosmology, Durham University, South Road, DH1 3LE, UK\label{aff87}
\and
Department of Physics and Astronomy, University of Aarhus, Ny Munkegade 120, DK-8000 Aarhus C, Denmark\label{aff88}
\and
Waterloo Centre for Astrophysics, University of Waterloo, Waterloo, Ontario N2L 3G1, Canada\label{aff89}
\and
Department of Physics and Astronomy, University of Waterloo, Waterloo, Ontario N2L 3G1, Canada\label{aff90}
\and
Perimeter Institute for Theoretical Physics, Waterloo, Ontario N2L 2Y5, Canada\label{aff91}
\and
Centre National d'Etudes Spatiales -- Centre spatial de Toulouse, 18 avenue Edouard Belin, 31401 Toulouse Cedex 9, France\label{aff92}
\and
Institute of Space Science, Str. Atomistilor, nr. 409 M\u{a}gurele, Ilfov, 077125, Romania\label{aff93}
\and
Departamento de Astrof\'isica, Universidad de La Laguna, 38206, La Laguna, Tenerife, Spain\label{aff94}
\and
Dipartimento di Fisica e Astronomia "G. Galilei", Universit\`a di Padova, Via Marzolo 8, 35131 Padova, Italy\label{aff95}
\and
Caltech/IPAC, 1200 E. California Blvd., Pasadena, CA 91125, USA\label{aff96}
\and
Institut f\"ur Theoretische Physik, University of Heidelberg, Philosophenweg 16, 69120 Heidelberg, Germany\label{aff97}
\and
Institut de Recherche en Astrophysique et Plan\'etologie (IRAP), Universit\'e de Toulouse, CNRS, UPS, CNES, 14 Av. Edouard Belin, 31400 Toulouse, France\label{aff98}
\and
Universit\'e St Joseph; Faculty of Sciences, Beirut, Lebanon\label{aff99}
\and
Departamento de F\'isica, FCFM, Universidad de Chile, Blanco Encalada 2008, Santiago, Chile\label{aff100}
\and
Universit\"at Innsbruck, Institut f\"ur Astro- und Teilchenphysik, Technikerstr. 25/8, 6020 Innsbruck, Austria\label{aff101}
\and
Satlantis, University Science Park, Sede Bld 48940, Leioa-Bilbao, Spain\label{aff102}
\and
Infrared Processing and Analysis Center, California Institute of Technology, Pasadena, CA 91125, USA\label{aff103}
\and
Instituto de Astrof\'isica e Ci\^encias do Espa\c{c}o, Faculdade de Ci\^encias, Universidade de Lisboa, Tapada da Ajuda, 1349-018 Lisboa, Portugal\label{aff104}
\and
INFN-Bologna, Via Irnerio 46, 40126 Bologna, Italy\label{aff105}
\and
INAF, Istituto di Radioastronomia, Via Piero Gobetti 101, 40129 Bologna, Italy\label{aff106}
\and
Astronomical Observatory of the Autonomous Region of the Aosta Valley (OAVdA), Loc. Lignan 39, I-11020, Nus (Aosta Valley), Italy\label{aff107}
\and
Center For Advanced Mathematical Sciences, American University of Beirut PO Box 11-0236, Riad El-Solh, Beirut 11097 2020, Lebanon\label{aff108}
\and
Junia, EPA department, 41 Bd Vauban, 59800 Lille, France\label{aff109}
\and
ICSC - Centro Nazionale di Ricerca in High Performance Computing, Big Data e Quantum Computing, Via Magnanelli 2, Bologna, Italy\label{aff110}
\and
Instituto de F\'isica Te\'orica UAM-CSIC, Campus de Cantoblanco, 28049 Madrid, Spain\label{aff111}
\and
CERCA/ISO, Department of Physics, Case Western Reserve University, 10900 Euclid Avenue, Cleveland, OH 44106, USA\label{aff112}
\and
Laboratoire Univers et Th\'eorie, Observatoire de Paris, Universit\'e PSL, Universit\'e Paris Cit\'e, CNRS, 92190 Meudon, France\label{aff113}
\and
IRFU, CEA, Universit\'e Paris-Saclay 91191 Gif-sur-Yvette Cedex, France\label{aff114}
\and
Dipartimento di Fisica e Scienze della Terra, Universit\`a degli Studi di Ferrara, Via Giuseppe Saragat 1, 44122 Ferrara, Italy\label{aff115}
\and
Istituto Nazionale di Fisica Nucleare, Sezione di Ferrara, Via Giuseppe Saragat 1, 44122 Ferrara, Italy\label{aff116}
\and
Universit\'e de Strasbourg, CNRS, Observatoire astronomique de Strasbourg, UMR 7550, 67000 Strasbourg, France\label{aff117}
\and
Kavli Institute for the Physics and Mathematics of the Universe (WPI), University of Tokyo, Kashiwa, Chiba 277-8583, Japan\label{aff118}
\and
Dipartimento di Fisica - Sezione di Astronomia, Universit\`a di Trieste, Via Tiepolo 11, 34131 Trieste, Italy\label{aff119}
\and
NASA Ames Research Center, Moffett Field, CA 94035, USA\label{aff120}
\and
Bay Area Environmental Research Institute, Moffett Field, California 94035, USA\label{aff121}
\and
Minnesota Institute for Astrophysics, University of Minnesota, 116 Church St SE, Minneapolis, MN 55455, USA\label{aff122}
\and
Institute Lorentz, Leiden University, Niels Bohrweg 2, 2333 CA Leiden, The Netherlands\label{aff123}
\and
Institute for Astronomy, University of Hawaii, 2680 Woodlawn Drive, Honolulu, HI 96822, USA\label{aff124}
\and
Department of Physics \& Astronomy, University of California Irvine, Irvine CA 92697, USA\label{aff125}
\and
Department of Astronomy \& Physics and Institute for Computational Astrophysics, Saint Mary's University, 923 Robie Street, Halifax, Nova Scotia, B3H 3C3, Canada\label{aff126}
\and
Departamento F\'isica Aplicada, Universidad Polit\'ecnica de Cartagena, Campus Muralla del Mar, 30202 Cartagena, Murcia, Spain\label{aff127}
\and
Department of Physics, Oxford University, Keble Road, Oxford OX1 3RH, UK\label{aff128}
\and
Institute of Cosmology and Gravitation, University of Portsmouth, Portsmouth PO1 3FX, UK\label{aff129}
\and
Department of Computer Science, Aalto University, PO Box 15400, Espoo, FI-00 076, Finland\label{aff130}
\and
Ruhr University Bochum, Faculty of Physics and Astronomy, Astronomical Institute (AIRUB), German Centre for Cosmological Lensing (GCCL), 44780 Bochum, Germany\label{aff131}
\and
DARK, Niels Bohr Institute, University of Copenhagen, Jagtvej 155, 2200 Copenhagen, Denmark\label{aff132}
\and
Universit\'e PSL, Observatoire de Paris, Sorbonne Universit\'e, CNRS, LERMA, 75014, Paris, France\label{aff133}
\and
Universit\'e Paris-Cit\'e, 5 Rue Thomas Mann, 75013, Paris, France\label{aff134}
\and
Department of Physics and Astronomy, Vesilinnantie 5, 20014 University of Turku, Finland\label{aff135}
\and
Serco for European Space Agency (ESA), Camino bajo del Castillo, s/n, Urbanizacion Villafranca del Castillo, Villanueva de la Ca\~nada, 28692 Madrid, Spain\label{aff136}
\and
ARC Centre of Excellence for Dark Matter Particle Physics, Melbourne, Australia\label{aff137}
\and
Centre for Astrophysics \& Supercomputing, Swinburne University of Technology,  Hawthorn, Victoria 3122, Australia\label{aff138}
\and
School of Physics and Astronomy, Queen Mary University of London, Mile End Road, London E1 4NS, UK\label{aff139}
\and
Department of Physics and Astronomy, University of the Western Cape, Bellville, Cape Town, 7535, South Africa\label{aff140}
\and
ICTP South American Institute for Fundamental Research, Instituto de F\'{\i}sica Te\'orica, Universidade Estadual Paulista, S\~ao Paulo, Brazil\label{aff141}
\and
Oskar Klein Centre for Cosmoparticle Physics, Department of Physics, Stockholm University, Stockholm, SE-106 91, Sweden\label{aff142}
\and
Astrophysics Group, Blackett Laboratory, Imperial College London, London SW7 2AZ, UK\label{aff143}
\and
INAF-Osservatorio Astrofisico di Arcetri, Largo E. Fermi 5, 50125, Firenze, Italy\label{aff144}
\and
Dipartimento di Fisica, Sapienza Universit\`a di Roma, Piazzale Aldo Moro 2, 00185 Roma, Italy\label{aff145}
\and
Centro de Astrof\'{\i}sica da Universidade do Porto, Rua das Estrelas, 4150-762 Porto, Portugal\label{aff146}
\and
HE Space for European Space Agency (ESA), Camino bajo del Castillo, s/n, Urbanizacion Villafranca del Castillo, Villanueva de la Ca\~nada, 28692 Madrid, Spain\label{aff147}
\and
Institute of Astronomy, University of Cambridge, Madingley Road, Cambridge CB3 0HA, UK\label{aff148}
\and
Department of Astrophysics, University of Zurich, Winterthurerstrasse 190, 8057 Zurich, Switzerland\label{aff149}
\and
Dipartimento di Fisica, Universit\`a degli studi di Genova, and INFN-Sezione di Genova, via Dodecaneso 33, 16146, Genova, Italy\label{aff150}
\and
Theoretical astrophysics, Department of Physics and Astronomy, Uppsala University, Box 515, 751 20 Uppsala, Sweden\label{aff151}
\and
ASTRON, the Netherlands Institute for Radio Astronomy, Postbus 2, 7990 AA, Dwingeloo, The Netherlands\label{aff152}
\and
Kapteyn Astronomical Institute, University of Groningen, PO Box 800, 9700 AV Groningen, The Netherlands\label{aff153}
\and
Anton Pannekoek Institute for Astronomy, University of Amsterdam, Postbus 94249, 1090 GE Amsterdam, The Netherlands\label{aff154}
\and
Department of Physics, Royal Holloway, University of London, TW20 0EX, UK\label{aff155}
\and
Department of Astrophysical Sciences, Peyton Hall, Princeton University, Princeton, NJ 08544, USA\label{aff156}
\and
Niels Bohr Institute, University of Copenhagen, Jagtvej 128, 2200 Copenhagen, Denmark\label{aff157}
\and
Center for Cosmology and Particle Physics, Department of Physics, New York University, New York, NY 10003, USA\label{aff158}
\and
Center for Computational Astrophysics, Flatiron Institute, 162 5th Avenue, 10010, New York, NY, USA\label{aff159}
\and
INFN-Sezione di Milano, Via Celoria 16, 20133 Milano, Italy\label{aff160}
\and
Departamento de F{\'\i}sica Fundamental. Universidad de Salamanca. Plaza de la Merced s/n. 37008 Salamanca, Spain\label{aff161}
\and
Thales~Services~S.A.S., 290 All\'ee du Lac, 31670 Lab\`ege, France\label{aff162}
\and
Observatorio Nacional, Rua General Jose Cristino, 77-Bairro Imperial de Sao Cristovao, Rio de Janeiro, 20921-400, Brazil\label{aff163}
\and
Center for Frontier Science, Chiba University, 1-33 Yayoi-cho, Inage-ku, Chiba 263-8522, Japan\label{aff164}
\and
Department of Physics, Graduate School of Science, Chiba University, 1-33 Yayoi-Cho, Inage-Ku, Chiba 263-8522, Japan\label{aff165}
\and
Institut de Physique Th\'eorique, CEA, CNRS, Universit\'e Paris-Saclay 91191 Gif-sur-Yvette Cedex, France\label{aff166}
\and
Sterrenkundig Observatorium, Universiteit Gent, Krijgslaan 281 S9, 9000 Gent, Belgium\label{aff167}
\and
Aurora Technology for European Space Agency (ESA), Camino bajo del Castillo, s/n, Urbanizacion Villafranca del Castillo, Villanueva de la Ca\~nada, 28692 Madrid, Spain\label{aff168}}    




%
%
   \abstract{
The near-infrared calibration unit (NI-CU) on board \Euclid's \ac{NISP} is the first astronomical calibration lamp based on \acp{LED} to be operated in space. 
\Euclid\ is a mission in ESA's Cosmic Vision 2015--2025 framework to explore the dark universe and provide a next-level characterisation of the nature of gravitation, dark matter, and dark energy.
Calibrating photometric and spectrometric measurements of galaxies to better than 1.5\% accuracy in a survey homogeneously mapping  $\sim$\,14\,000\,deg$^2$ of extragalactic sky requires a very detailed characterisation of \ac{NIR} detector properties as well as constant monitoring of them in flight.
To cover two of the main contributions -- relative pixel-to-pixel sensitivity and non-linearity characteristics -- and to support other calibration activities, NI-CU was designed to provide spatially approximately homogeneous ($<$\,12\% variations) and temporally stable illumination (0.1\%--0.2\% over 1200\,s) over the NISP detector plane with minimal power consumption and energy dissipation. NI-CU covers the spectral range $\sim$\,[900,1900]\,nm -- at cryo-operating temperature -- at five fixed independent wavelengths to capture wavelength-dependent behaviour of the detectors, with fluence over a dynamic range of $\gtrsim$\,100 from $\sim$\,15\,ph\,s$^{-1}$\,pixel$^{-1}$ to $>$\,1500\,ph\,s$^{-1}$\,pixel$^{-1}$. 
For this functionality, NI-CU is based on \acp{LED}. We describe the rationale behind the decision and design process, the challenges in sourcing the right \acp{LED}, and the qualification process and lessons learned. We also provide a description of the completed NI-CU, its capabilities, and performance as well as its limits. NI-CU has been integrated into \ac{NISP} and the \Euclid\ satellite, and since \Euclid's launch in July 2023, it has started supporting survey operations.

}
%
%
%
%
   \titlerunning{\Euclid. IV. The NISP Calibration Unit}
   \authorrunning{Euclid Collaboration: F.\ Hormuth et al.}
   
   \maketitle
%
%
%
%
\section{Introduction}\label{sec:Intro}

ESA's \Euclid\ mission is setting out to provide a much better understanding of the nature of gravitation, dark matter, and dark energy \citep{Laureijs11,euclid2024}. 
\Euclid will employ a number of probes, weak lensing, baryon acoustic oscillations, and other tools to measure the expansion history of the Universe and to map the structure formation across cosmic time \citep{euclid2024,Amendola18} in a celestial survey covering $\sim$\,14\,000\,deg$^2$, more than a third of the total sky \citep{Scaramella-EP1}. 
By observing structural and spectral information of more than one billion galaxies over this area in a survey over six years, \Euclid\ will provide a major next step in cosmology and produce an unrivalled database for structure and photometry of cosmic objects in the visible and \acf{NIR} that is unlikely to be surpassed, at least in the next three decades \citep{Scaramella-EP1}.

To reach the required surface number density of galaxies, shape measurements are needed for galaxies to a brightness of 24.5\,mag (AB, 10$\sigma$ point sources) in the visible, with redshifts being derived down to point-source limits of at least 24.0\,mag (AB) in the NIR, in order to be combined with matched external ground-based photometry at shorter wavelengths. A similar-level \ac{NIR} photometric sensitivity requirement arose from baryon acoustic oscillation galaxy clustering science. To extract accurate spectroscopic redshifts for galaxy clustering, NIR spectroscopy needs to reach a flux limit $\le$\,2$\times10^{-16}$\,erg\,cm$^{-2}$\,s$^{-1}$ at $3.5\sigma$ for any emission line at 1600\,nm from a \ang{;;0.5} diameter source.

\Euclid\ was designed to carry a payload that is capable of delivering this. The spacecraft consists of a 1.20\,m telescope aperture feeding three instrumental modes: a wide-field visible wavelength high-fidelity imaging instrument \citep[VIS;][]{cropper2018,VIS2024}, as well as a multi-passband photometer and a slitless spectrometer -- combined into the Near-Infrared Spectrometer and Photometer \citep[NISP;][]{Maciaszek22,NISP2024}. NISP will provide both spectroscopic redshifts directly, and contribute the required \ac{NIR} passbands to photometric redshifts.

All cosmological probes require not only large numbers of galaxies, but also extremely accurate data about galaxy shapes, as well as distances. Therefore the observed data have to be photometrically and spectrophotometrically calibrated in order to provide the required accuracy in colours, spectral energy distributions, or derived quantities such as redshifts. In this paper we describe the background, design, manufacturing, testing, and the actual performance of the near-infrared calibration unit (NI-CU) inside NISP. NI-CU has the function to facilitate calibration and monitoring of the 16 \ac{NIR} detectors in NISP's focal plane, with respect to their relative pixel-to-pixel sensitivity variations (i,e.\ `small-scale flatfield'), non-linear pixel response, and other aspects.

In this article we aim at providing crucial reference and background information for astronomers and engineers alike, describing the motivations for NI-CU's design its capabilities and limits. Following a brief primer on \Euclid and an intro to NISP, we describe the underlying requirements driving NI-CU's design (Sect.~\ref{sec:design}), the process of finding suitable LEDs, and the challenges involved in using them for the first time in a space-based astronomical calibration lamp (Sect.~\ref{sec:leds}). We follow this with the derived final NI-CU design (Sect.~\ref{sec:nicu-design}), information about manufacturing and testing (Sect.~\ref{sec:manufacturing}), and the performance and limits of the resulting flight hardware (Sect.~\ref{sec:performance}). We close with Sect.~\ref{sec:lessons}, we reflect on the whole development process and collect lessons learned, both from a technical as well as a managerial point of view, and provide thoughts on potential improvements that other projects working on a similar LED-based calibration approach could consider.

\section{NISP and its calibration unit NI-CU: requirements and design}
\label{sec:design}

\subsection{The NISP instrument: design and calibration approach}
\label{sec:NISP}

Following \Euclid's goals, the resulting instrumental requirement specifications and driver for the subsequent design of NISP were two-fold:

\bi
\item The galaxy clustering probe required spectroscopy-based redshift measurements of emission-line galaxies out to redshift $z\sim2$, that is in the NIR at spectral resolution $R\gtrsim 250$ for 1\arcsec\ sized galaxies. For this, slitless NIR spectroscopy was found to be the most efficient approach to cover large areas on the sky with a high multiplexing factor, using grisms with different relative orientation angles to reduce confusion of line identification for overlapping adjacent objects. 
\item As part of the weak-lensing probe, photometric redshifts over the survey area of 14\,000\,deg$^2$ and out to redshifts $z\sim2$ required a combination of deep (ground-based) visible light passbands and three NIR passbands between $\sim$\,0.95 and 2\,\micron\ that required being observed from space to reach the brightness limit and area in finite time.
\ei

The combination of these probes into one NIR instrument set the basic design principle of changeable optical elements in a single light path towards a single detector system as well as observations in parallel with VIS, whose light would be reflected off with a dichroic beam-splitter before entering NISP. The details of NISP's layout and characteristics are laid out in \citet{maciaszek2016,Maciaszek22} and \citet{NISP2024}, but in order to elucidate the calibration source design decisions discussed below, a short summary is needed.

NISP uses a lens-based optic \citep{grupp2012,NISP2024} to focus the light beam entering the instrument from the dichroic element through a passband filter or a grism onto the detector array.
It is required to provide three changeable passband filters and two types of grisms, one of which is to be observed in multiple orientations of dispersion direction. As a result, NISP contains a rotatable filter wheel assembly (FWA) with five positions; three filters \YE, \JE, and \HE\ \citep{Schirmer-EP18}; and a light-tight closed position for dark exposures. The fifth position is open to transmit light towards the grism wheel assembly (GWA), also with five positions. Aside from an open position to let light from the imaging mode pass, the wheel contains one `blue' grism (BG$_\mathrm{E}$) and three versions of the `red' grism (RG$_\mathrm{E}$) in different orientations \citep{costille2019,Maciaszek22}. Both the FWA and GWA are encased in a light-tight enclosure. A corrector and a camera optic before and after the wheels are responsible for properly imaging the light onto the focal plane.

\begin{figure}[ht]
\begin{center}
\includegraphics[angle=0,width=\columnwidth]{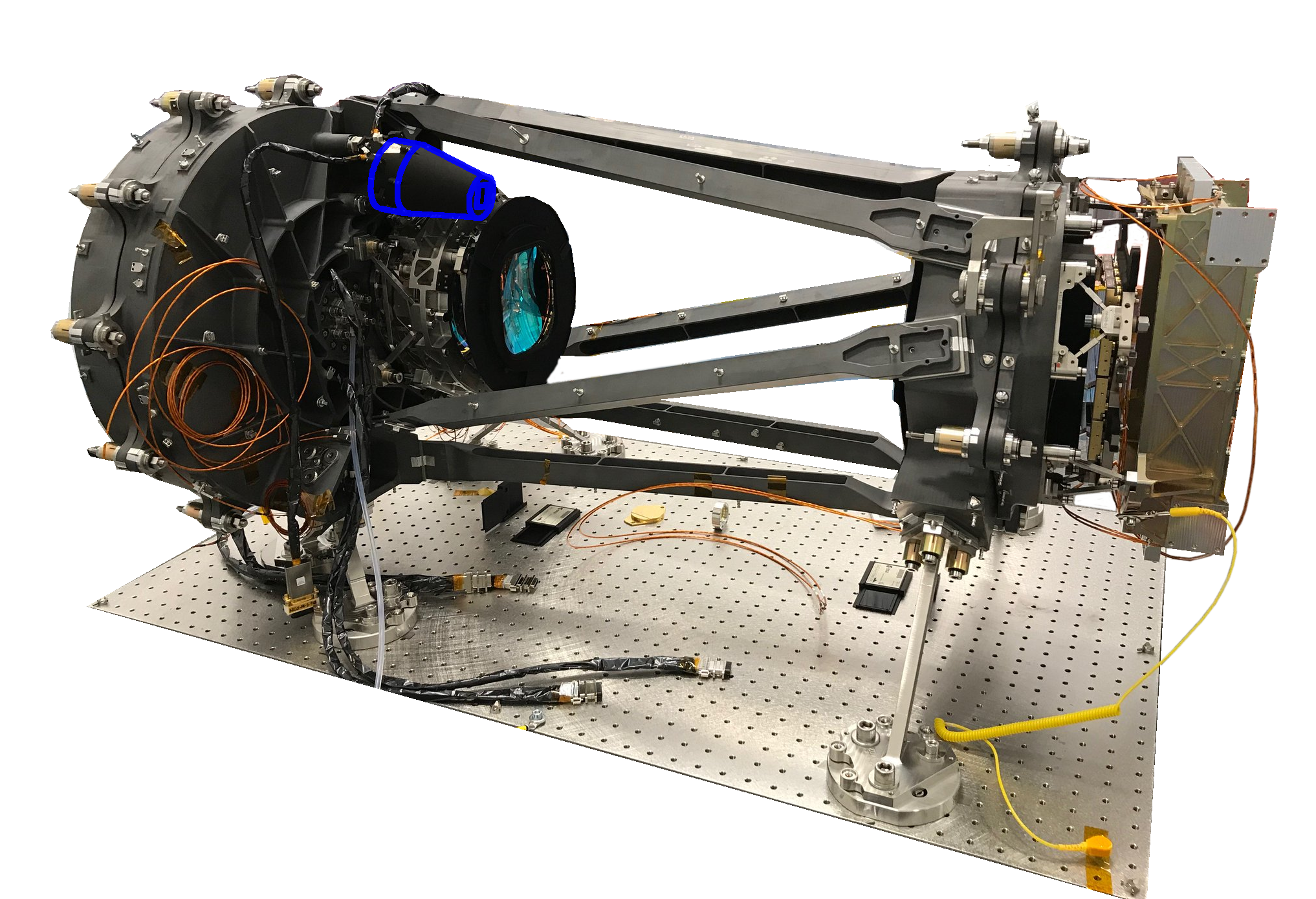}
\end{center}
\caption{NISP flight model before wrapping in light-tight multi-layer insulation. The left side contains the optics with encased filter- and grism-wheel assembly, the right side the detector array and read-out electronics. The black-coated NI-CU calibration unit, mounted just next to the optics, has been outlined in blue for better visibility.}
\label{fig:NISP_cutout}
\end{figure}

NISP's \ac{FPA} consists of 16 Hawaii2RG detectors \citep{secroun2016} in a 4$\times$4 grid. The FPA is located at around $\sim$\,700\,mm distance from the last lens. Importantly, both optics and FPA are operating at cryo-temperatures, $\sim$\,135\,K and $\sim$\,95\,K, respectively. These temperatures are reached by passive cooling through radiators on the spacecraft, to remove all heat that is generated in the detectors and their read-out electronics on one side, and the wheel motors and the calibration light source on the other. 

Together with some heaters, these elements provide a rather stable optical and electronic system that, jointly with a survey that provides an only weakly varying orientation angle towards the Sun, can rely on operating in a narrow temperature range.
\medskip

Apart from a required high sensitivity of the FPA detectors, the top-level requirement is that of a 1.5\,\% relative photometric accuracy for NISP photometric observations -- the core requirement driving the design of the NISP calibration source NI-CU, as described in the following. In the case of vanishing Poisson shot noise -- of hypothetical astronomical infinite signal-to-noise sources on the sky -- the derived photometric flux of such a source across all regions of the survey and over time should not vary more than 1.5\,\%. This includes every aspect of spatial and temporal variations of the filter passband, potentially effects of degrading optics, but also all properties of the detectors: contributions of a baseline bias, dark current, charge persistence, non-linearity, brighter-fatter-effect, crosstalk, and spatial and temporal variations of the quantum efficiency (QE), that is the flat-field. Since all these contributions and spatial and temporal variations are or at least could be non-zero, they have to be characterised, and calibrated across the survey duration. The overall calibration framework for \Euclid\ instruments is rather complex and will be described in a comprehensive review in the future. A description of all NISP components, NISP performance, and initial overview over calibration can be found in \citet{NISP2024}.

\subsection{NI-CU requirements: origin and description}

To reach the functionalities above, a driving aspect for NI-CU was on one side to calibrate the pixel-to-pixel flatfield of the 16 detectors, meaning the relative QE or sensitivity of all pixels as a function of wavelength.  The second requirement was to provide light for linearity calibration: Hawaii2RG and similar detectors are intrinsically non-linear in their charge to signal relation. In the NISP detector operating mode of `multi-accumulate' up-the-ramp sampling this appears when the increasing amount of charge in a pixel creates a deviation from a constant slope in mapping of incidence photons to read-out countrate at the low and/or high end. This was to be calibrated with the calibration source as well. Both were supposed to be achieved by combining NI-CU in-flight use with data from NI-CU and other measurements in the lab before flight.

Initial design considerations of NI-CU covered different options regarding its location in the NISP optical train, with implications for its scope, capabilities, and resulting technical challenges:
At some point it became clear that it would not be practical or even possible to illuminate the detectors along the optical path of the instrument in order to shine light also through the NIR-filters. Impractical, because any light-source would neither have a flat intensity over the whole wavelength range nor an intensity distribution resembling any particular celestial source. In any case there was no compelling reason to include a light-path through the filters, since the goal was to calibrate the time-variable characteristics of the detectors, not the filters that were expected to experience near-zero changes over their lifetime, even in a harsh space environment.
An illumination of the detector system along NISP's optical path was also seen as almost impossible, as it would have either led to vignetting of the science beam by the calibration source, or would have required the use of a mechanism to move the calibration source into and out of the beam for calibration use, adding major complexity and risk.\footnote{The whole NISP instrument has only two mechanisms, each one rotating the filter- and grism-wheel.} An early version of the calibration source up to development phase A ($\sim$2011) still had such an approach in place, including fibres illuminated by integration spheres that contained tungsten lamps as light sources. This optical design was discarded due to its complexity, while LEDs as alternatives to tungsten filaments were considered early on.

The scientific requirements for the NISP detector calibration were broken down into corresponding functional and performance requirements for NI-CU. The primary requirement was to provide relative and small-scale flatfield calibration of the detector array, therefore excluding any effects related to the telescope or NISP instrument optics. `Relative' here indicates that any kind of absolute photometric calibration of the instrument was not within its scope,\footnote{Absolute photometric calibration of detectors just from an on-board lamp is extremely challenging as it is fundamentally a `chicken-and-egg' problem: any such system using an internal lamp would have to also calibrate the lamp's behaviour over time, which in turn requires a calibration system that demonstrably does not drift in sensitivity itself (see discussion in Sect.~\ref{sec:abscal}). As is the case for NISP, most missions hence combine relative sensitivity calibration with externally-calibrated photometric standard objects in the sky, observed during the operations, to put an instrument's sensitivity on an absolute scale.} and `small-scale' describes that pixel-to-pixel sensitivity variations over a few 100 pixels distance are to be calibrated, not large scale changes over for instance a full detector. The latter is handled by a `self-calibration' approach using survey data.

The handover scale of $\sim$\,100 pixels was motivated by the view that with a constant illumination neighbouring pixels would see the same number of photons -- within the shot noise -- and hence their resulting signal levels would then be a measure of relative sensitivity. A perfectly `flat' illumination over the full \ac{FPA} would obviously take care of all pixels at the same time, but this is very challenging to achieve, both from artificial light sources or celestial references. Hence the pixel-to-pixel sensitivity variations were enabled by the assumption of only locally constant illumination levels over $\sim$\,100 pixel patches, with the large-scale tie in delegated to use of celestial objects. This scale was both in reach for a calibration light source, while not being not too small for the use of self-calibration.

It was further required to illuminate the detector at five discrete, narrow-band wavelengths and at five different illumination levels ranging from $4.3\times10^{10}$\,ph\,s$^{-1}$\,m$^{-2}$ to at least $4.3\times10^{11}$\,ph\,s$^{-1}$\,m$^{-2}$ in the detector plane, corresponding to $\sim$\,15--150\,ph\,s$^{-1}$\,pixel$^{-1}$,\footnote{These were the original requirements, with a dynamic range of 10. The design later converged towards the ability for also substantially higher fluences, and the resulting dynamic range is now being used in calibration.} with an accuracy and reproducibility of 10\%. This was to be achieved at least at any time during the nominal 6.5 years of mission duration. The reproducibility was mainly meant to be able to predict and properly command count rates on the detectors in flight, not as an absolute calibration requirement.
Actual wavelengths did not have to correspond to the bandpasses of the NISP filters, but were to cover the NISP wavelength range of 900--2000\,nm in a more or less evenly spaced manner to enable interpolation of calibration data to any desired wavelength and hence passband. The target bandwidth of each illumination channel was required to be in the range of 10--200\,nm in order to both have clear separation of at least two LED wavelength bands per NISP filter, while avoiding any potential effects from just monochromatic light. This effectively ruled out the use of laser diode type emitters -- too narrow emission -- or tungsten filaments -- too wide emission -- without additional bandpass filters. 

To reduce straylight and therefore the potential for difficult to predict inhomogeneous illumination of the detector, the maximum permitted light intensity at any point outside of the detector area was restricted to 10\% of the average flux inside the detector area. Together with an appropriate design of the calibration source's illumination and baffeling approach this was required in order to achieve a maximum peak-to-valley large-scale illumination non-uniformity of $\sim$\,10\% over the full detector array -- for reasons of homogeneous signal-to-noise in calibration exposures -- while assuming that large-scale throughput variations for example by the optics are properly characterised by the self-calibration approach.

To achieve the goal of small-scale flatfield calibration, spatial uniformity was required to be better than 0.1\% ultimately defined to apply for any 1\,mm scale ($\sim$\,55 pixels) in the detector plane, rendering the design sensitive to caustics or specular reflections on for instance sharp edges. The small-scale homogeneity would make sure that by design no such sharp illumination features should be present, permitting the assumption that pixels located close to each other would indeed see the same illumination.

Temporal stability requirements for the generated illumination levels applied after a grace period or warmup-time of 180s after setting the required flux level. These were specified as a maximum linear drift of 0.2\% over a 1200\,s period, and a maximum RMS fluctuation of 0.3\% after removal of the linear drift component. This was mainly motivated by the typical timescales on which calibration sequences were to be performed.

It was not required to operate more than one wavelength channel at the same time. A non-activated channel was restricted to emit less than 500\,ph\,s$^{-1}$ at any time due to driver leakage currents or pickup of electromagnetic interference. Assuming isotropic emission over one hemisphere this corresponds to less than 0.001\,ph\,s$^{-1}$ per pixel over the course of a 100\,s calibration exposure, that is a fully negligible yet technically feasible level.
Each channel was required to fulfil all these requirements for a total operation time of more than 800 hours, including more than 30\,000 changes of the illumination level, operating at typically 130--140\,K and after having survived the mechanical loads of the launch and transfer to the final satellite position.
As a last requirement, the calibration unit had to be able to provide sufficient diagnostic data or signals to assess its system health.

\subsection{NI-CU design decisions}

Based on these requirements, three major initial design decisions were taken at the start of NI-CU development:
\bi 
\item Location of NI-CU (position in instrument, relative orientation to detector),
\item illumination principle (optics, straylight avoidance), and
\item type of light source (narrow-band vs. broadband source with additional bandpass filters) and control principle.
\ei 

The location of NI-CU had to be chosen at a suitable distance from the detector plane in order to meet the large-scale uniformity requirement at a reasonable technical effort. At the same time any obstruction of the nominal light path of the instrument itself was not permitted. As the finished NISP instrument is enshrouded in a black Kapton envelope, the only available position was inside the instrument on the so-called panel P1, next to the optical corrector-lens assembly \citep[see][]{Maciaszek22}. This location provided an in principle unobstructed off-axis view\footnote{In the NISP FM there is indeed a small stripe on one side of the FPA of order 30 pixels width, and another small path on the opposite side seeing vignetting due to the late addition of an FPA-baffle.} of the full detector array at a distance of $\sim$\,60\,cm. The resulting location of NI-CU in the assembled NISP flight model is shown in Fig.~\ref{fig:NISP_cutout}.

With a direct view of the detector, the decision on the illumination principle centred on whether to use one or more point-like sources. Limited space and requirements on robustness made the use of integrating spheres and optical fibres less appealing. In addition, the large-scale uniformity requirement of $\sim$\,10\% proved difficult to meet with a simple point source at the given distance from the FPA and array size. A solution that combined simplicity, small volume and sufficient mixing of all wavelength into a single point of origin while meeting the flux level requirement was found in using a small piece of optical diffuser material with a Lambertian scattering profile. An optimised and innovative choice of size and orientation of this diffuser is a central part of the design work -- this is described in detail in Sect.~\ref{sec:nicu-design}.

Regarding the light-source, the minimum full width at half maximum (FWHM) requirement of 10\,nm for each wavelength channel ruled out the use of laser diodes as emitters. For tungsten filaments the upper limit of 200\,nm would have required adding optical filters, but not fundametally ruling them out. LEDs were considered from an early stage of the project, likely to meet wavelength criteria without additional effort.
The tradeoff between tungsten filaments (i.e.\ traditional light bulbs) and LEDs was basically about heritage, ageing, and wavelength stability:

{\bf Heritage:} Tungsten lamps have been used in space for decades, for instance in ISO's ISOCAM \citep{cesarsky1996}, HST's NICMOS \citep{schultz2003} and WFC3/IR \citep{khandrika2021}, JWST's MIRI \citep{wright2015} and NIRSpec \citep{jakobsen2022}. Tungsten will also still be standard for upcoming ground-based instrumentation, such as ESO E-ELT MICADO \citep{rodeghiero2020}. Infrared LEDs on the other hand have little astronomy space heritage as light source, other than as part of optocouplers with limited wavelength coverage and limited performance requirements. There are a few examples of LEDs used for calibration in Earth-observation experiments, for example the Indian RESOURCESAT satellites \citep{resourcesat}, or the DESIS instrument \citep{desis2019} on the International Space Station, with LEDs at wavelength below $\sim$\,1\,\micron\ -- but heritage remains limited.

{\bf Ageing:} While tungsten lamps suffer from ageing as the filaments become thinner with increasing time of use, they are not sustaining radiation damage as LED semiconductors do. The effect of ageing on tungsten filaments will be a gradual change of filament resistance and light output -- and ageing can likely be predicted quite well. The performance of LEDs can, in principle, suffer both due to continuous radiation effects and isolated space weather events such as coronal mass ejections. Sufficient testing and application of substantial margins were thought to mitigate this issue.

{\bf Wavelength stability:} The main challenge for tungsten filaments is the requirement to produce at least five different illumination levels differing by at least one order of magnitude: using only one filament for all flux levels could only be achieved by changing the average electrical power and hence filament temperature. A change in temperature results in a change of the spectrum, shifting to shorter wavelengths at higher power and temperature. Therefore, this would require use of a relatively narrow bandpass filter to achieve independence between flux level and wavelength, and hence a low overall efficiency. Using more than one tungsten lamp per wavelength channel could achieve different flux levels with similar spectral properties, at the cost of substantially higher complexity, control effort, and component count.

LEDs also have a temperature dependency of the wavelength -- inverse to tungsten, with a blueshift at colder temperatures -- but the shift between room temperature and operational conditions can be measured on the ground and is very predictable. The comparably low power consumption of LEDs leads to very low internal heating, and hence the resulting temperature changes during operations are mostly negligible. A change of the output intensity at practically constant wavelength characteristic can easily be achieved over two or more decades by adjusting the drive current or applying digitally controlled pulses with modulation to a constant current source. No additional filters are required to produce the desired wavelengths, which allows for a very simple and efficient system, since all of the generated light is emitted in-band and put to its purpose instead of being largely absorbed. 

Initial tests about market availability and radiation sensitivity (Sect.~\ref{sec:leds}) led us to choose the LED-approach. The decision was based on the excellent stability over the required time scale of 1200\,s given a stable drive-current, good controllability of the output flux with good efficiency and hence low power dissipation, a simple design, as well as the opportunity to actually produce a novel heritage of using infrared LEDs as a light sources in space.

\section{NI-CU LED selection} \label{sec:leds}
\subsection{Initial candidate screening}
The first step of NI-CU development was identifying and initial screening of candidate LED devices to prove feasibility of an LED-based design and as a foundation for the subsequent formal space qualification.
Selected LEDs had to fulfil the main requirements explained above (wavelength distribution, spectral and flux stability) and to prove their reliability in a cryovacuum environment.
As explained, no suitable LEDs with sufficient space heritage existed, nor were LED suppliers catering to the needs of a space hardware development project in terms of specific technical qualification aspects,  as well as manufacturer quality and procurement standards.
LEDs in the wavelength range up to 1550\,nm at room temperature are relatively abundant, but market research showed that for longer wavelengths only one manufacturer, LMSNT,  seemed to exist.\footnote{LEDs manufactured by LMSNT ({\em LED Microsensor NT}, based in Russia) were sold by multiple suppliers under different names, but similarities between properties and datasheets traced them to a single manu\-facturer.}

For initial downselection, a wide range of off-the-shelf LEDs was procured from a range of manufacturers -- at least at shorter wavelengths -- and in a range of packages. These included plastic packages, hermetic TO (`transistor outline') packages, nominally hermetic packages with integrated reflectors, and reflector packages with direct access to the LED die. Wavelengths ranged from 970 to 2000\,nm at room temperature. 
An initial optical inspection, to assess overall workmanship of the devices, and initial operation at cryogenic temperatures removed the first LED package types -- LEDs with a silicone glob-top, meaning silicone plastics covering the LED semiconductor, are not suitable for cryogenic operation. The silicone will become rigid and crack at low temperatures, creating mechanical stresses on the LED die, and resulting in severe reduction of LED performance or even LED failure.
Excluding these LED types still left at least one type per wavelength channel available for further evaluation of basic characteristics.

The laboratory equipment for the initial tests was centred on a small temperature-controlled cryostat based on a closed-cycle cooler, equipped with ample electrical connections and two large diameter glass fibre ports. In combination with external source measurement units for LED control and a NIR spectrometer as well as an optical power meter, this allowed flexible experimental setups with reasonable turn-around times. High-power heaters inside the cryostat enabled fast warm-up times, while the small cryostat volume and mass resulted in relatively short evacuation and cooling times.
The use of external and internal fibres for light measurements enabled simple spectral and flux measurements. Flux stability measurements were limited by the dependency of the setup's transmission on external parameters such as the laboratory temperature or mechanical stress on the fibres. For improved flux measurements, especially at low levels, an internal extended InGaAs-photodiode and transimpedance amplifier circuit were mounted on the cryostat's cold-plate close to the LED under investigation. 

Laboratory screening experiments at temperatures as low as 100\,K confirmed that all LED candidates worked under such conditions and produced stable output. Measurements of the dependency of electrical parameters, optical flux, central wavelength, and FWHM of the LED output on the operating temperature were performed at this stage. This initial screening was carried out with two spectrometers ranging from 500--2500\,nm to also check for potential emission outside the LED's nominal wavelength range. Simple temperature shock experiments with liquid nitrogen also confirmed a high degree of resilience, surprising at least in the case of the plastic packages. 

Subsequent tests at cryogenic temperatures confirmed the stability of the LED emission. Long-term measurements of the optical output over a range of 60\,000 activation cycles with an on-off ratio 1:4 did not show any failures. Flux stability always satisfied the requirements on the 1200\,s time scale or for the whole experiment duration.
These tests showed that the major part of LED `instability' occurs during the first seconds after initial application of drive current. Internal heating of the LED will reduce its light output and lead to a noticeable flux decrease during the first seconds. This stops only once thermal equilibrium is reached. Meeting the stabilisation time requirement of 180\,s hence implied a proper thermal design of the LED mount with a thermally conductive path to a heatsink or cold baseplate.

Operation at cold temperatures were found to be as one could expect for LEDs based on standard semiconductor physics: 

\bi
\item An increase of the forward voltage of the p-n junction with decreasing temperature,
\item a corresponding higher electrical input power when driven with constant current, and hence a higher optical output power,
\item a further increase of optical output power vs.\ electrical input power due to a higher efficiency of the LEDs at cold temperatures,
\item a shift of the emitted light toward shorter wavelengths,
\item a decrease of the FWHM of the emitted light, and
\item a highly stable output, once sufficient time is allowed for stabilisation,
\item no emission outside the narrow nominal wavelength range.
\ei

An early radiation experiment was performed with these LEDs to assess the impact of irradiation with high-energy protons. The dominant effect of proton radiation\footnote{Ionising photon radiation was not seen as a source of possible defects, as one would only expect transient effects at the radiation levels typically experienced in NISP.} -- the main components of cosmic rays during the mission -- on LEDs is a persistent decrease in light output efficiency due to displacement damage, requiring higher electrical power to achieve the same light output as before the irradiation occurred. 
We irradiated a total number of 25 LEDs of five different types and wavelengths -- at warm conditions and with the LEDs not operated -- with an 10\,MeV equivalent proton dose of $9\times10^9$\,cm$^{-2}$, using the proton beam mode of the HIT Heidelberg facility.\footnote{Heavy Ion beam Therapy facility of the Heidelberg University Hospital, Germany} This corresponds to twice the expected lifetime dose of the \Euclid mission at the planned NI-CU location.
Electrical and optical parameters of the LEDs were measured before and ca.\ 30\,min after ending the irradiation, at room temperature. Depending on the LED type -- and hence the involved semiconductor material -- a decrease of the LED light output between 0\% and 48\% was observed. This confirmed that the chosen LED candidates could be used in the expected radiation environment without danger of a total failure, given that sufficient margins for the optical flux were applied. It also underlines the fact that LEDs designed for different wavelengths -- and therefore based on different materials -- can suffer from very different grades of radiation damage. Here measurements are required, and estimates or extrapolations will not be sufficient. 

These measurements and tests narrowed down the electrical and optical parameters of the LEDs over the full range of operating conditions and allowed us to define the preliminary electrical interface for NISP's control electronics (instrument control unit, ICU). 
Measurement of the actual forward current and voltage was identified as the means to monitor LED health and determine the actual electrical input power at any time.
The technical realisation of the LED driver by the ICU was to be based on a modulated constant current source with a 4-wire connection to each NI-CU LED. The current source itself was not replicated for each LED channel but multiplexed, so that only one LED at any time could be operated.

\subsection{Final selection, procurement, and qualification}

We settled on five devices manufactured by the companies Epigap (Japan) and LMSNT (Russia) for in-depth characterisation and formal space qualification. Basic manufacturer-provided parameters of these LEDs are given in Table~\ref{table:ledspecs}, for different current levels and operating modes, at room temperature. While the Epigap devices are specified for up to 100\,mA DC current (for `CW', continuous wave operation, i.e.\ 100\% duty cycle), LMSNT defines the `Quasi-CW' mode, that is operation at 50\% duty cycle with a rectangular signal at up to 200\,mA peak current.

We defined common drive parameters for all LED types to unify the electrical interface and to apply proper de-rating according to ECSS\footnote{European Cooperation for Space Standardization,\newline \href{https://ecss.nl/standards}{https://ecss.nl/standards}} requirements: A core current range of 10--100\,mA, pulse-width-modulated (PWM) with a rectangular waveform with duty cycles between 5 and 50\%. The resulting total dynamic range of 100 provided sufficient margin to match the requirement and accommodate for performance degradation of the LEDs during the mission lifetime.\footnote{To go beyond a dynamic range of order 100 likely requires the use of multiple sets of LEDs and e.g.\ nested integration spheres as in the case of the {\it Roman} Space Telescope. See Sect.~\ref{sec:dynamicrange} for a discussion.} PWM duty cycles shorter than 5\% were not foreseen in order to avoid non-linearities, while lower currents down to 1\,mA are possible, though with an increased output drift, since the NI-CU power supply in the flight hardware would guarantee 0.1\% stability at 10\,mA, but proportionally lower stability at lower currents.

Wavelength and FWHM of the LMS20LED-C device (channel E) are clearly larger than the required values for NI-CU at room temperature, emphasising the need for cryogenic testing.

\begin{table*}[htb]
\caption{NISP final LED types and parameters.}
\label{table:ledspecs}
\centering
\begin{tabular}{c c c c c r@{}l c}
\hline\hline\noalign{\vskip 1pt}
Channel & Manufacturer & Model & $\lambda_\mathrm{peak}$   & FWHM  &  \multicolumn{2}{c}{Output} & Material\\
 &  & & [nm] & [nm]  & \multicolumn{2}{c}{power} \\

\hline\noalign{\vskip 1pt}
A & Epigap & EOLC-970-17    & 970   &  70        &   2\,mW @& 20\,mA CW   & GaAs\\
B & Epigap & EOLC-1200-17-1 & 1200  &  70        &   3.5\,mW @& 100\,mA CW& InGaAsP/InP\\
C & Epigap & EOLC-1550-17-1 & 1550  &  130       &   2.5\,mW @& 100\,mA CW& InGaAsP/InP\\
D & LMSNT  & LMS18LED-C     & 1845  &  100--200  &   0.7--1.1\,mW @& 200mA Quasi-CW& GaInAsSb/AlGaAsSb\\
E & LMSNT  & LMS20LED-C     & 2045  &  150--250  &   0.8--1.2\,mW @& 200mA Quasi-CW& GaInAsSb/AlGaAsSb\\
\hline

\end{tabular}
\tablefoot{
Wavelength, width, and power output are given here for room temperature. Resulting values at cold operating temperatures are given in Table~\ref{tbl:ledcold}. Output power are representative values provided by the manufacturer at 100\% `continuous wave' duty cycle for the Epigap LEDs, and 50\% duty cycle for LEDs D and E. LED A is made from a single material, LEDs B--E from two-material heterostructures.
}
\end{table*}

Findings regarding packaging quality, especially the die attach between semiconductor and header base of the LMSNT devices, led to the decision to procure the LEDs as unpackaged dice and organise packaging through a dedicated service supplier, resulting in a uniform manufacturing process and material selection, and in a much better control of the whole packaging process and quality. As a side result we achieved mechanical similarity between all LEDs, leading to a more simplified and interchangeable LED mount design for the final unit.

\begin{figure}[ht]
\begin{center}
\includegraphics[angle=0,width=0.7\columnwidth]{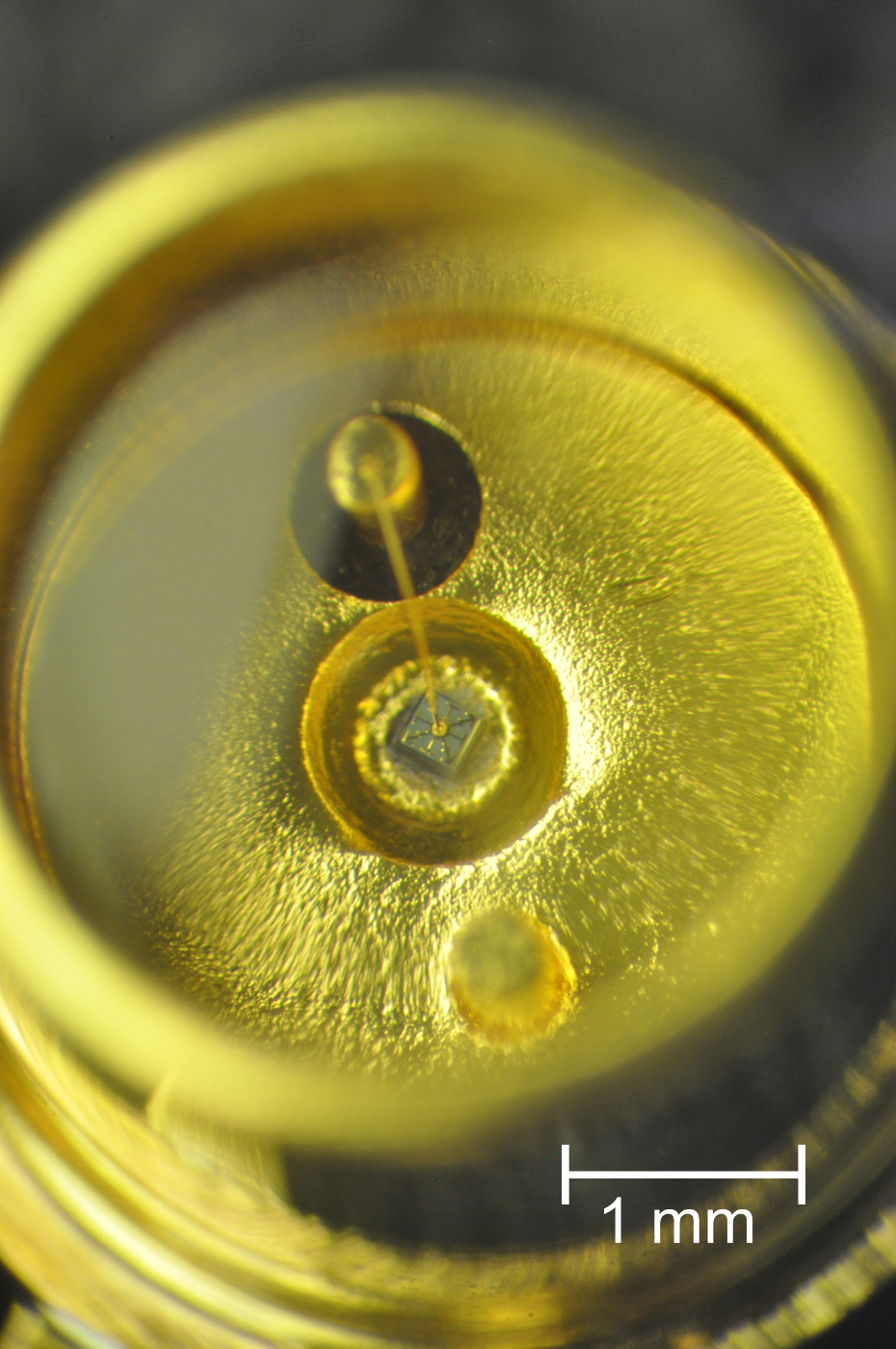}
\end{center}
\caption{Closeup of finished LED: wire-bonded semiconductor die (centre), inside gold-coated housing, behind hermetically sealed glass (image courtesy \textsl{von Hoerner \& Sulger}).}
\label{fig:NICU_LED}
\end{figure}

All five LED types were mounted on standard Kovar two-pin TO-18 headers manufactured by Technotron, Germany and hermetically sealed with a Kovar cap with glass window manufactured by Schott, Germany. Assembly was conducted by First Sensor Lewicki, Germany. Die attach was performed with EPO-TEK EJ2189 silver-filled epoxy glue, and the second electrical connection was realised by ball bonding with 25\,\micron\ Au wire. Sealing was done via seam welding, followed by gross and fine leak testing, serialisation, and a simple electrical test before delivery for further testing. Apart from transient problems with the bonding process no significant problems were observed.
A close-up photograph of an NI-CU LED is shown in Fig.~\ref{fig:NICU_LED}.

To mitigate potential risks with procurement or significant quality differences between different batches, we decided to procure for each LED type enough devices from the same batch in order to be able to carry out all qualification and manufacturing activities plus have a sufficient margin in case of mishaps or defects. This implied setting up a mature plan for qualification and manufacturing relatively early during the project and resulted in the procurement of 200 LEDs of each type.

Formal space qualification commenced after homogeneous packaging of all devices and consisted mainly of the following steps:\footnote{Final LED procurement oversight, space qualification, mechanical design, as well as manufacture, assembly and tests of the NI-CU STM, EM, FM, and FS models were carried out by {\em von Hoerner \& Sulger}, Schwetzingen, Germany. Substantial part of the LED qualification steps were carried out by {\em RoodMicrotec}, Nördlingen, Germany.}

\bi 
\item Cryogenic electrical and optical test, lifetime test, and subsequent inspection and definition of electrical burn-in parameters for flight devices, performed on subgroups.
\item  Parallel long-term cold storage test of a further subgroup, followed by thorough inspection for failures or degradations.
\item  Thermal cycling, electrical testing, PIND\footnote{Particle Impact Noise Detection -- a standard test to detect loose particles inside semiconductor packages} and X-ray inspection, and optical measurement of 60 devices per type, from which final flight devices were to be selected (`screened devices').
\item Setting aside 20 screened devices per type as final flight LEDs to be used in the deliverable NI-CU models, including contingency.
\item Thorough analysis and testing of the remaining 40 screened devices per type, including: extensive non-destructive and destructive analysis both before and after thermal cycling, mechanical tests (vibrational loads), radiation dose testing in warm conditions (for a discussion see Sect.~\ref{subsec:techlessons}), humidity tests and cryogenic optical and electrical characterisation. 
\ei

Many of these tests followed the requirements laid out by the applicable ECSS (see above) and ESCC\footnote{European Space Components Coordination,\newline \href{https://spacecomponents.org}{https://spacecomponents.org}} standards are a common practice for non-space semiconductors as well. Particularly in the detailed destructive and X-ray analysis of the LEDs we received valuable support by the French space agency CNES and, through them, by Thales Alenia Space, Toulouse.

Final tests concerned the susceptibility of the -- highly efficient -- LEDs to unwanted photon emission due to small possible current leakage in the driver electronics, or pickup of electromagnetic interference via the harness routed inside the satellite. First investigation were carried out in our test cryostat where the LED was driven by a high-frequency (HF) current equivalent to what would be expected due to electromagnetic interference coupling into the harness. They showed that it was needed to control the forward voltage resulting from the worst case expected leakage current to a value well below the typical forward voltages at which photon emission occurred. This was achieved by adding a 100\,k$\Omega$ in parallel connection to the driver output on the warm electronics side, which reduced bias voltages due to leakage currents to well below 100\,mV where LEDs do not yet operate, while having a negligible effect on voltage vs.\ current characteristics and output linearity.

Actual measurements with additional HF currents did not show any significant unwanted emissions over the frequency range of interest as the combined forward voltage due to leakage and electromagnetic interference did not reach the excitation threshold. At high frequencies this is further mitigated due to the relatively high junction capacitances of the LEDs used. Operation at cold temperatures further helps the case as the forward voltage and hence the excitation threshold increases. 

This was confirmed by qualification tests in a more complex cryogenic setup, targeting the electromagnetic compatibility of the overall cold detector electronics in the context of instrument development. This experiment was performed in a cryostat with a Delrin cover, permitting irradiation of a flight representative piece of NI-CU harness with HF signals from the outside. The tested LEDs were positioned in front of the detector, creating a very sensitive measurement of any spurious emission. This experiment did confirm conformance to the requirements as well.\footnote{Though for NI-CU we did not encounter any issues with unwanted emissions, this is a real risk if not managed and tested properly. The effects of leakage currents and their worst case magnitude over the full mission lifetime have to be taken into account. Otherwise, the shutter-less calibration source would add background noise for each and every science frame.}

The overall qualification campaign certified that the selected LEDs and manufacturing steps are capable to withstand the harsh environmental conditions of the \Euclid mission and to the space environment in general. While performance degradations due to radiation damage were observed as expected, no failure or excessive loss of output flux was encountered. 
Defects or failures due to mechanical or thermal loads were not observed. Electrical and optical parameters at different temperatures were found to be predictable and well-behaved.
The electro-optical measurements of the screened devices at operational temperatures around 135\,K permitted to accurately predict wavelength characteristics and electrical parameters of the flight devices at cold operating temperatures. A summary is given in Table~\ref{tbl:ledcold}.

\begin{table*}[htb]
\caption{NISP final LED parameters at 135\,K.} 
\label{tbl:ledcold}
\centering
\begin{tabular}{c c c c c}
\hline\hline\noalign{\vskip 1pt}
    Channel    & $\lambda_\mathrm{peak}$ & FWHM &  Wavelength shift & Forward voltage \\
 & [nm]             & [nm]   &  [nm/K]  & [V] @ 100\,mA, 50\,\% \\
\hline\noalign{\vskip 1pt}
A & \phantom{0}939  &  \phantom{0}24  & 0.12 & 2.23\\
B & 1157 &  \phantom{0}46  & 0.14 & 1.09 \\
C & 1467 &  \phantom{0}82  & 0.14 & 0.96 \\
D & 1735 &  129 & 0.57 & 1.68 \\
E & 1873 &  \phantom{0}92  & 0.58 & 0.93 \\
\hline
\end{tabular}
\tablefoot{
For each LED channel the peak wavelength and width are listed, as well as the shift in central wavelength with temperature in the regime around 135\,K. The last column contains the characteristic forward voltage at 100\,mA drive current and 50\% PWM that can be used as a direct diagnostic of LED brightness, and that defines the maximal power consumption.
}
\end{table*}

The selection, test, and qualification phase of the NI-CU LEDs ended successfully with 20 screened LEDs per wavelength channel with full formal ESA-accepted space qualification and sufficient knowledge of key performance parameters at the intended operational conditions.

\section{Final NI-CU design}
\label{sec:nicu-design}

With LED selection and qualification running in parallel, the mechanical and optical design of NI-CU was developed along these
main requirements:
\bi
\item Illumination concept compatible to an angled and off-axis optical path towards the detector;
\item limited build volume due to proximity to the NISP optics;
\item interface to the instruments SiC panel is sensitive to shear stress, requiring reduction of forces in the interface plane over the full non-operational temperature range;
\item generation of an almost flat illumination pattern with sharp transition from full illumination to almost zero light intensity at the edges of the detector plane;
\item combination of five different wavelengths with proper output levels;
\item full redundancy by providing two LEDs per wavelength channel;
\item low impact on the instrument's thermal budget;
\item good short-term stability and reproducibility with quick stabilisation time of the output intensity after activation;
\item protection of any sensitive components inside NI-CU to survive long-term storage on the ground;
\item electrical driver in NISP's overall warm electronics for a centralised commanding and power supply \citep[see][]{NISP2024}.
\ei

\begin{figure*}[htb]
\begin{center}
\includegraphics[angle=0,width=0.9\textwidth]{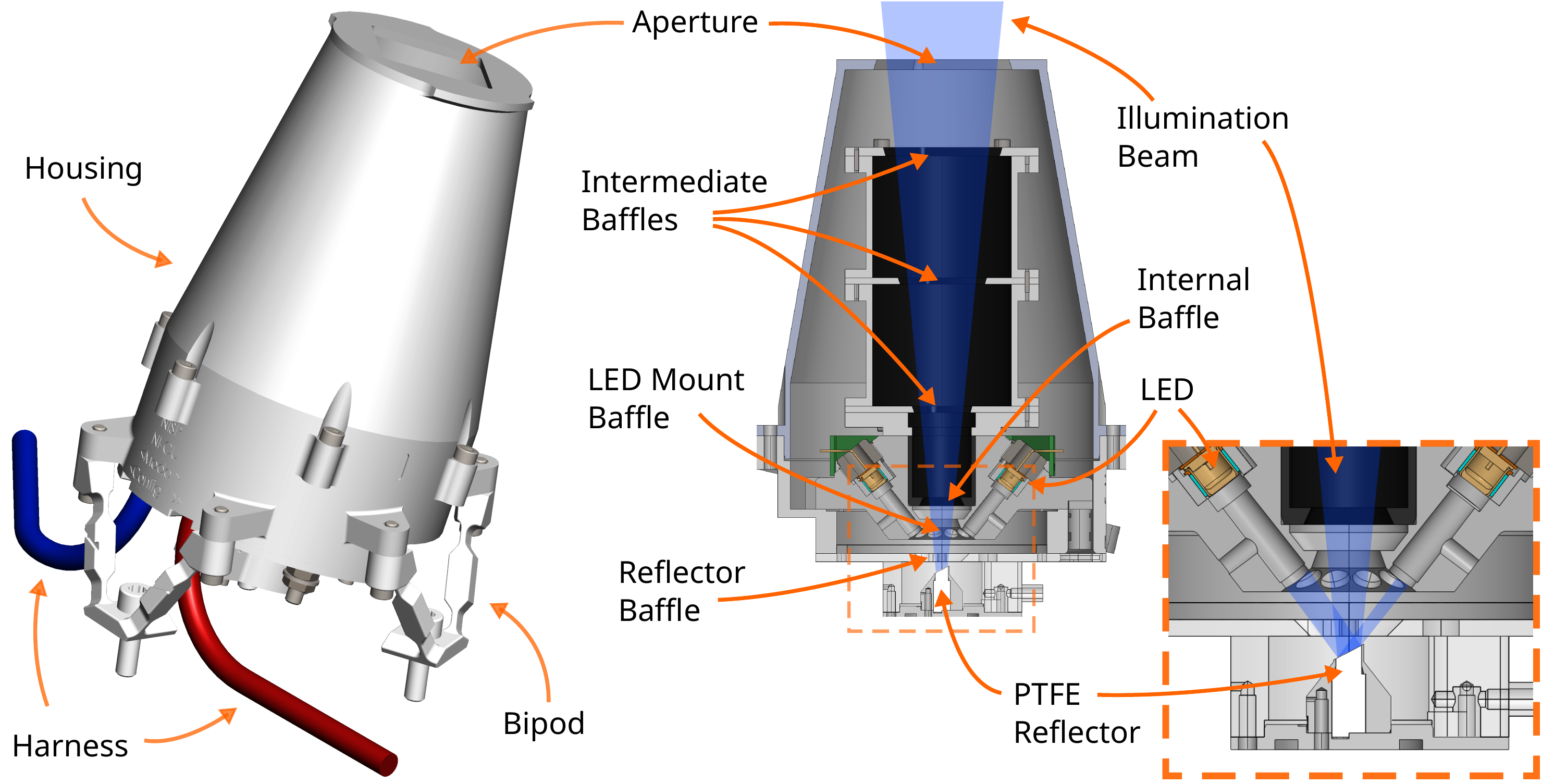}
\end{center}
\caption{Final NI-CU design. {\em Left:} Outer view of housing, bipod interface to NISP, and harness locations. {\em Centre:} Cross-section of the NI-CU main body with illumination-critical components. The LEDs are pointing downward, illuminating the reflector patch with tilted surface (enlarged on the right). A number of baffles inside NI-CU shapes the beam to just illuminate the detector array, while minimising straylight.}
\label{fig:nicu-outer}
\end{figure*}

The mechanical design can be roughly divided into few main components as shown in Fig.~\ref{fig:nicu-outer}. The first main part is the NI-CU housing. It contains the LEDs and electrical wiring, and performs light mixing and shaping to generate the required illumination pattern with proper shape, uniformity and level at the focal plane. It is machined of aluminium and coated to a large extent with {\em PNC} black coating\footnote{ PNC is a black coating manufactured by MAP, France, \url{https://www.map-coatings.com}, having both a high solar absorptance ($\alpha_\mathrm{S}=0.97$) and emissivity ($\epsilon=0.91$).} to reduce unwanted straylight and out-of-field illumination as much as possible. Light leaves the housing through an aperture on the top end and is directed to the NISP focal plane.

The second main component comprises three short titanium bipods which connect the NI-CU body to the interface points on the P1 SiC panel. These bipods provide sufficient stiffness to ensure sufficient pointing accuracy of NI-CU over the applicable temperature range, but are flexible enough to absorb stress related to mechanical loads caused by vibration and acceleration as well as by thermal coefficient mismatches of the employed materials during cool-down of the instrument. This is especially important as the feet of these bipods are bolted to dedicated Invar interface pads on the instrument's P1 panel, which are in turn glued to to underlying SiC structure. This glue interface is prone to potential failures due to shear stress, which could be induced by thermal stress. The design was optimised in this respect to rule out a detachment of NI-CU even in the case of the loss of one bipod or glue connection.

Two separate electrical harnesses -- one for the primary set of LEDs and one for the secondary set of LEDs -- leave the bottom of the unit and are routed on the instrument's SiC structure to a connector bracket.

The cross-section drawing of the optical elements in the right panel of Fig.~\ref{fig:nicu-outer} shows the internal layout. The bottom part of the aluminium housing accommodates the LEDs and a small diffusor manufactured from Spectralon which produces a near-homogeneous illumination of the detector plane, similar for each LED.
Spectralon is a sintered PTFE (Polytetrafluoroethylene) type material with excellent near-Lambertian reflectance properties over the NI-CU LED wavelength range. The Lambertian properties result in a smooth illumination pattern without high-frequent spatial intensity variations, which is almost independent of the actual illumination pattern produced by the individual LEDs.
The light cone reflected by the Spectralon target is further shaped by a set of three intermediate baffles and exits the unit through its final aperture. The baffles use custom designed asymmetric shapes in order to produce a rectangular field of illumination at the detector plane despite the off-axis layout of the system. Internal edges of the baffles and the apertures are tapered to avoid grazing reflections.

The LEDs are mounted above the diffusor into aluminium nitride bushings for the purpose of electrical insulation and good thermal connection to the NI-CU structure at the same time, helping to achieve the required stability criteria and short stabilisation times of the output flux. The LEDs point downward to the target, which is realised as a small pin made of Spectralon, encased in a metal jacket to expose only the top surface to the LED light.
The design option to optimise photon flux levels and to harmonise levels between the different LED types by using gold coated light concentrator tubes attached to the LEDs for additional beam shaping was not used in the end after final light-level measurements. With this design, in principle alignment of flux levels can be achieved by adjusting the length of the reflecting tubes. If necessary, further adjustments to match flux levels of vastly different types of LEDs could be made by using different coatings for individual concentrator tubes with different reflectivities.

\begin{figure}[ht]
\begin{center}
\includegraphics[angle=0,width=\columnwidth]{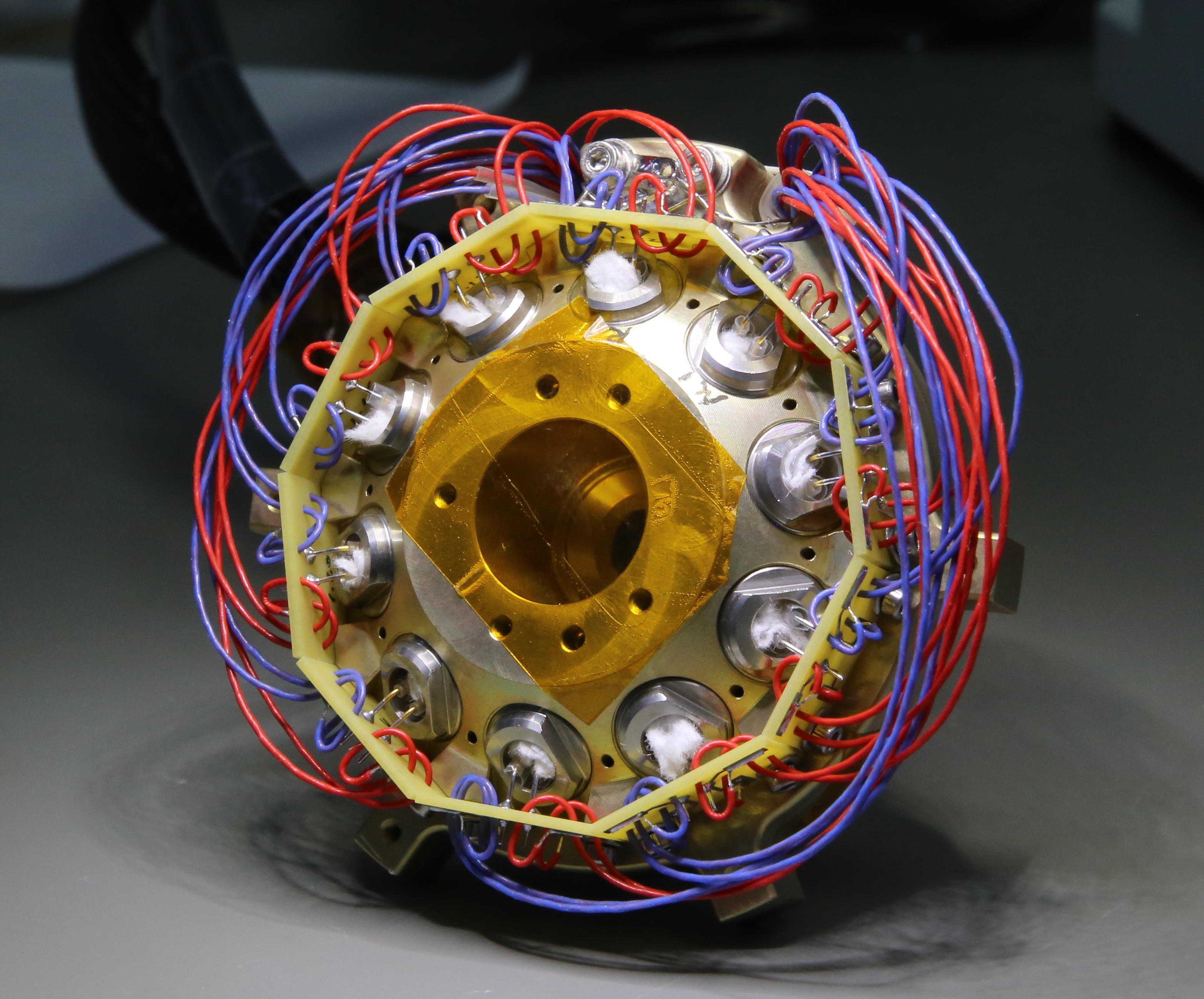}
\end{center}
\caption{View of the cable connections of the inward-pointing LEDs at the base of the NI-CU FM. The LEDs are looking inwards and are electrically connected to the harness cables via small individual printed circuit boards which provide a mechanical support against the impact of vibrations.}
\label{fig:NICU_EQM_base}
\end{figure}

A closer view of the LED and reflector arrangement is provided in the right panel of Fig.~\ref{fig:nicu-outer}.
The region around the reflector consists on the one hand of a reflective bowl concentrator above the Spectralon surface to increase efficiency, and on the other hand of a black cavity below reflector level. The latter serves to suppress straylight, that is to avoid light reaching the detector plane which does not originate from the Spectralon diffusor.
The LEDs are soldered to small printed circuit boards which also serve as connection point for the harness wires.
This assembly is shown in Fig.~\ref{fig:NICU_EQM_base}, looking towards the back sides of the LEDs with the full wiring of the 4-wire connections for control and measurement.

\begin{figure*}[ht]
\begin{center}
\includegraphics[angle=0,width=0.75\textwidth]{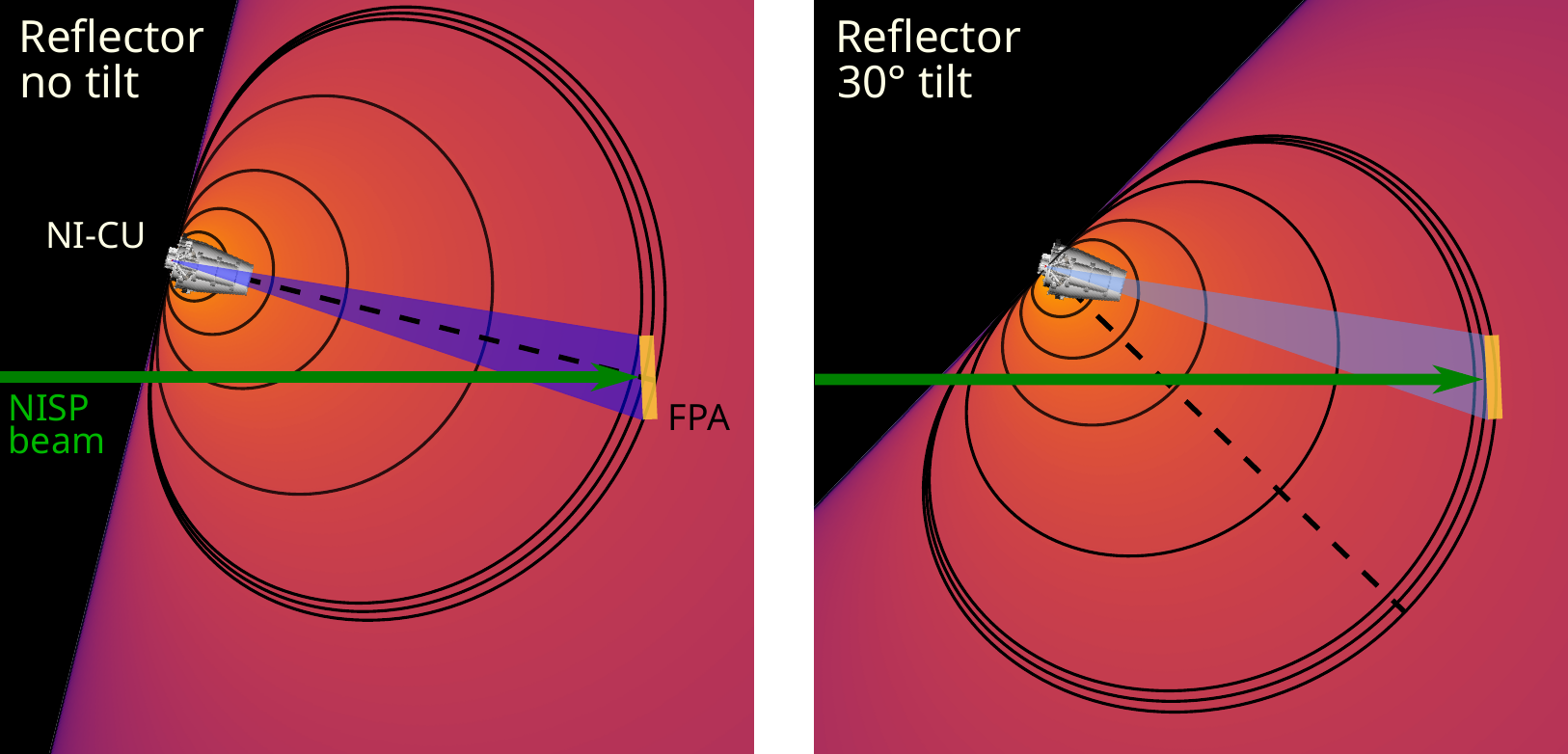}
\end{center}
\caption{Illustration of reaching a homogeneous illumination despite an off-centre position of NI-CU and a tilt angle between optical axis and FPA. Shown are the optical beam (green arrow), the slightly tilted FPA (yellow box), as well as NI-CU and the intensity of its illumination as colours and contours. The centre of the NI-CU emission coil is marked as a dashed line, the actual shaped beam is shown in blue.
{\em Left:} Flat reflector patch; the beam centre has the same orientation as the centre of the Lambertian reflection cone and is creating a substantial gradient across the tilted FPA.
{\em Right:} 30$^\circ$ tilted reflector patch; the effect of 30$^\circ$ rotation along the cosine component of the reflection cone matches the angle of NI-CU relative to the FPA, creating a near-homogeneous illumination.
}
\label{fig:NICU_illumination}
\end{figure*}

A particular design feature is visible upon a close look at the Spectralon diffusor surface in Fig.~\ref{fig:nicu-outer} (right panel): it is not perpendicular to the central axis of the unit, but tilted by $\approx$\,30$\degree$. This tilt angle is the result of an optimisation to provide illumination homogeneity  given the off-axis location of NI-CU and a tilt of the focal plane away from the optical axis:\footnote{\Euclid has a `Korsch' off-axis design that extends to its instruments.}
if the normal of the reflector surface would point directly towards the detector plane, a strong illumination-gradient would be present across the FPA (see Fig.~\ref{fig:NICU_illumination}). In the given geometry of the NISP instrument this would not allow us to achieve the required large-scale uniformity of $\lesssim$\,10\%. A tilt of the reflector, however, shifts the illumination along the cosine component of the Lambertian reflection law, which can be optimised -- with carefully chosen tilt angle and direction -- to match all tilt angles, resulting in a much more uniform light distribution.

A classic trade-off by simulation was performed in order to arrive at a good compromise between flux levels and homogeneity: The tilt angle changes how much projected Spectralon reflector surface is seen by different LED positions in NI-CU. The fact that in such a geometry the light of each LED hits the Spectralon target at a different angle opens up an opportunity to even out flux levels between different LED types by choosing the most appropriate location: the LED with the smallest flux at a given current can be placed in a location where the projected reflector surface is at its maximum, while the brightest LED (highest flux at same current) can be put into the opposite position. This feature therefore also allows us to optimise the use of the available dynamic range of the current source.

The diameter of the Spectralon target has also been subject to simulation and optimisation. A large target would lead to overall increased flux levels, but also to higher out-of-field illumination at the detector plane: the size of the transition region between the nominal field of illumination and the out-of-field area results from simple geometrical relations between the reflector size, its distance from the defining, beam-shaping aperture, and its distance from the detector plane. An oversized reflector would lead to very soft edges of the illumination pattern which in turn could provoke reflections or straylight originating from structures around the focal plane.

A very small reflector patch on the other hand would create a very sharp transition area, but decrease the achievable flux levels. It would also reduce the impact of the reflector tilt on the large-scale homogeneity. This becomes obvious if one imagines an infinitely small reflector path (i.e.\ a point), where the Lambertian reflector properties vanish and are replaced by a simple $1/r^2$ law insensitive to any tilt angles -- obviously a point does not have a surface to which a tilt angle can be applied.

\begin{figure*}[ht]
\begin{minipage}{\textwidth}
\centering
\raisebox{-0.5\height}{
\includegraphics[angle=0,width=0.5\textwidth]{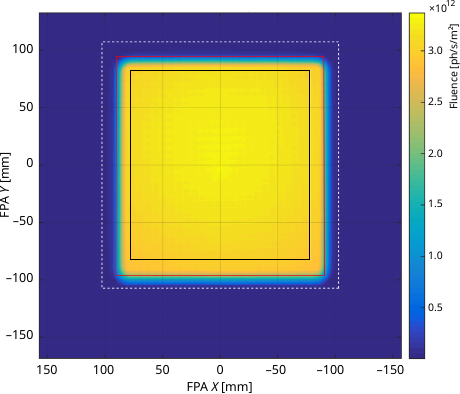}
}
\qquad
\raisebox{-0.5\height}{
\includegraphics[angle=0,width=0.40\textwidth]{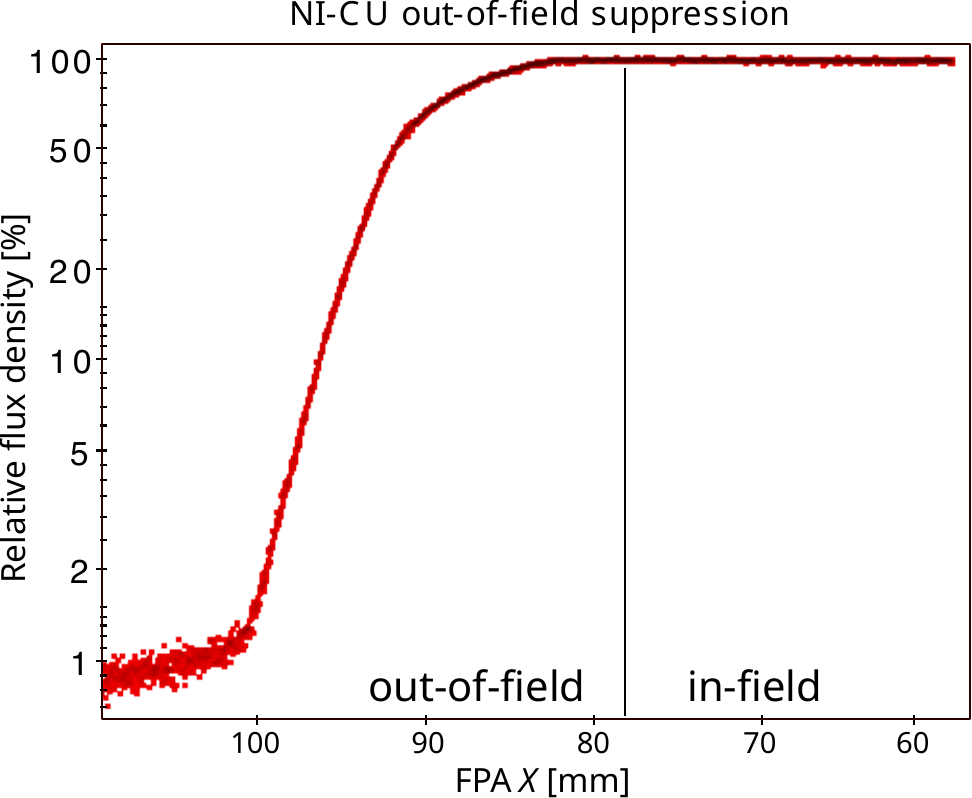}
}
\end{minipage}
\caption{NI-CU in-field and out-of-field illumination. {\em Left:} Reconstructed NI-CU EQM illumination in the focal plane using a test system without the full NISP. The fluence drops quickly outside the nominal in-field (black line) until it is close to zero 25\,mm further out (red line). The overall light emitted to outside the nominal field, when integrated out to the dashed line, is substantially below the requirement of 10\% of the total in-field light. {\em Right:} Closeup on the radial sharp transition in relative flux density at the edge of the nominal NI-CU illuminated field, here independently measured for the NI-CU VM -- the approximate edge of the nominal NISP FPA in-field is marked by the vertical line. This indicates a functioning baffling approach and creates a very low amount of straylight inside the NISP cavity from light falling outside the target FPA area.}
\label{fig:NICU_suppression}
\end{figure*}

As in the case of the tilt angle, optimisation was performed by simulation, taking expected LED flux levels into account. A reflector diameter of 3\,mm was chosen for the final design. Figure~\ref{fig:NICU_suppression} shows the measured dropoff in physical units as it would appear at the edge of the FPA. Within $\sim$\,25\,mm the illumination decreases by two orders of magnitude, minimising the amount of straylight in the instrument cavity.

One should keep in mind that the Lambertian reflector properties on which this design and the simulation results rely also depend on the thickness and the surface quality of the reflector. For a good reflector performance light must be able to penetrate the surface, as the scattering and diffusing process takes place inside the material and not just at the surface.
Regarding surface quality, the reflector must absolutely be protected from mechanical damage or contamination. As PTFE is a lipophilic material, contamination from oil or grease, even in small amounts, must be avoided. Such contamination would degrade both flatness of the reflectivity versus wavelength and the Lambertian properties with a tendency to add a specular reflection component.

In addition to appropriate handling and storage conditions during assembly and testing of the NI-CU units, a one-way purge valve was added to the reflector cavity, which enabled to flush the unit with dry nitrogen when required. 
A corresponding removable outlet valve can be attached to the exit aperture to maintain the nitrogen atmosphere for extended time periods, specifically during shipping or storage. Mounting of the outlet valve was designed in a way to prevent damage to the PNC black coating of outer unit surfaces visible from the location of detector plane and to keep the contact surface with the valve assembly at a minimum.

\section{NI-CU manufacturing and unit-level tests}\label{sec:manufacturing}

Following the model philosophy of the NISP instrument, five models of the NI-CU calibration unit were built and individually tested before delivery to the NISP project, namely a structural and thermal model (STM) and engineering and qualification model (EQM), a flight model (FM), a flight spare (FS), and an avionics verification model (AVM). This philosophy allowed the development team to clearly separate between models ready to fly in the actual mission (FM, FS) and models used primarily for test and qualification purposes (STM, EQM, AVM). Testing of the FM and FS was conducted at less stringent acceptance levels -- especially with respect to thermal and vibrational loads -- to avoid any overstressing prior to flight, while the STM and EQM were exposed to higher qualification levels. 

\begin{figure}[ht]
\begin{center}
\includegraphics[angle=0,width=\columnwidth]{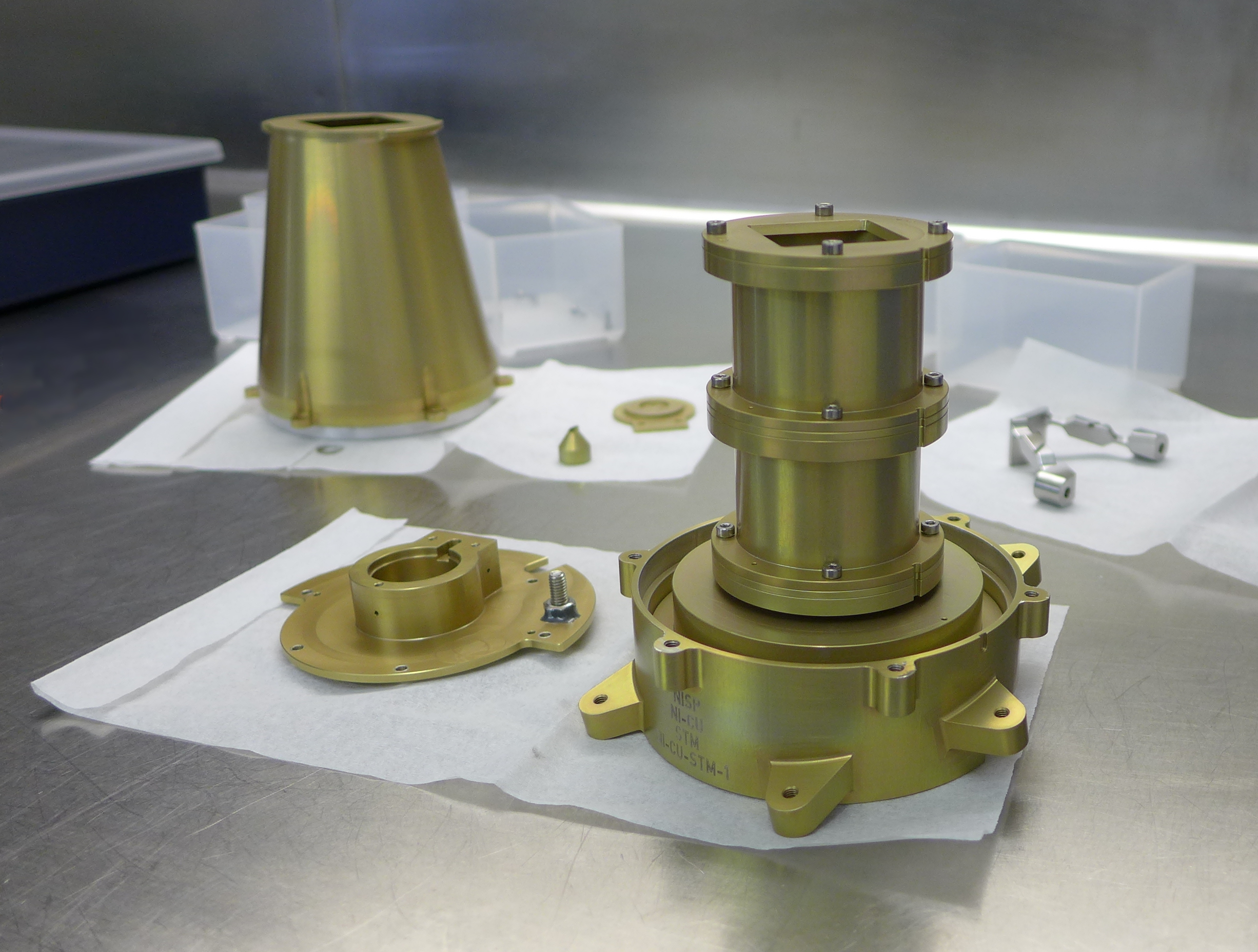}
\end{center}
\caption{NI-CU structural and thermal model before assembly, showing the inner structure and baffle cascade. For the flight model these components are painted with black PNC coating.}
\label{fig:NISP_STM}
\end{figure}

The NI-CU STM can best be described as a dummy with representative mechanical properties. This model was used to test structural integrity of the mechanical design on unit level under thermal and sinusoidal as well as random vibrational loads. It was further integrated into the STM of the NISP instrument, serving both as a mechanical interface check and to contribute to the representativeness of the NISP STM. The NI-CU STM was a fully passive unit without black coating, and without any LEDs or equivalent heat sources -- at a dissipated peak power of $\lesssim$\,110\,mW at any time, the thermal loads imposed by NI-CU on NISP were deemed negligible.
Building the STM allowed us to refine the design and manufacturing process. This experience helped to speed up later manufacturing of EQM, FM, and FS.
Parts of the STM during the assembly phase are shown in Fig.~\ref{fig:NISP_STM}.

The EQM was a fully functional unit with the same build standard as FM and FS.
The EQM not only served for final checks of the design according to qualification levels in the thermal and mechanical domain, but was also used to get first-hand measurements of actual photon flux at the NISP focal plane versus electrical drive parameters for all five LED channels.

In order to enable detailed control of the NI-CU EQM during test campaigns independently of the availability of the NISP ICU, a dedicated NI-CU test electronics (NI-CUT) were built and delivered. This unit allowed us remote control of all nominal and redundant LED channels, monitoring of actual LED currents and voltages, and measurement of output flux at the NI-CU exit aperture via a removable photodiode. The NI-CUT proved to be a valuable tool not just for the EQM campaign but throughout development of the NI-CU units themselves.

\begin{figure}[ht]
\begin{center}
\includegraphics[angle=90,width=0.7\columnwidth]{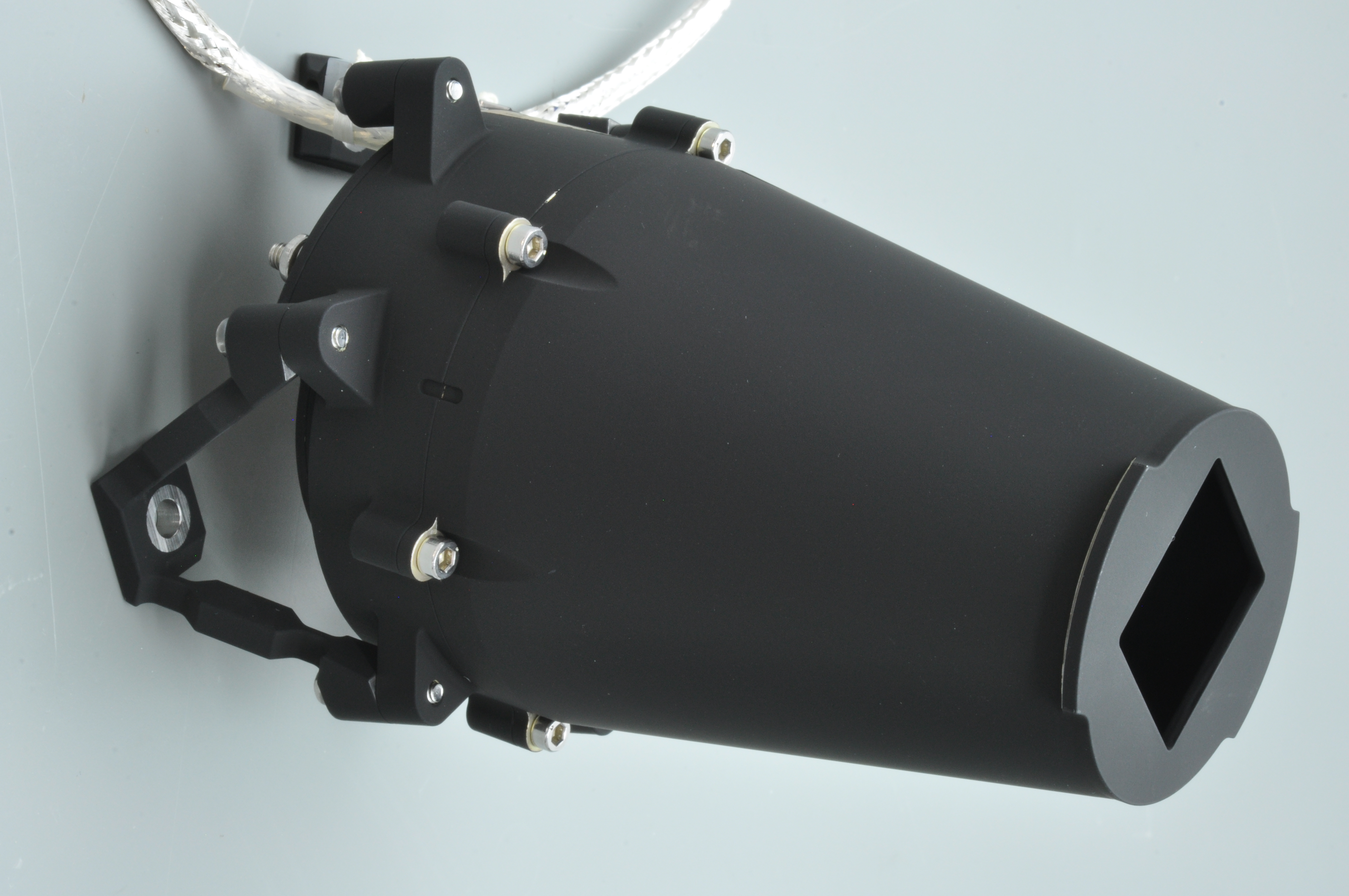}
\end{center}
\caption{NI-CU flight model before integration into NISP. The beginning of the $\sim$680\,mm nominal and redundant harnesses are visible at the bottom left, connecting to a long harness to the warm electronics at a connector bracket on the NISP structure. Despite being located far outside NISP's optical beam, it is fully coated with black PNC. Only screws and harness remain uncoated.}
\label{fig:NICU_FM}
\end{figure}

The flight models (FM and FS) were built with the single purpose of integration into the NISP flight model, and a potential spare, respectively. The finalised NI-CU flight model without protective cover is shown in Fig.~\ref{fig:NICU_FM}.

The models described above share a similar design, and especially their mechanical interface allows them to be integrated into the NISP instrument. This is not really necessary for tests or debugging activities with respect to the ICU or general telecommanding and housekeeping tasks. For this purpose an avionics model of NISP was built, including an AVM of the NI-CU calibration unit. The NI-CU AVM had to mimic only the electrical interface of the calibration unit, ideally for the whole range of operational temperatures, while being operated only at room temperature. The resulting need for a wide range of forward voltages was fulfilled by various combinations of diodes and visible LEDs. As a result the NI-CU AVM enabled to test the LED driver and housekeeping part of the ICU over the whole specified range of currents, voltages, and duty cycles, while giving some visible feedback for simple troubleshooting.

While the STM only underwent basic mechanical and thermal tests, the three fully functional EQM, FM, and FS units were also subjected to verifications of the achieved illumination strength and to different degrees also shape, uniformity and stability. 
Proper validation of the illumination pattern with the required signal-to-noise ratio turned out to be a challenging task: at the designed range of output levels laboratory measurements of infrared radiation at the design distance of the detector plane were rather infeasible. 

A further NI-CU model was therefore built, the so-called validation model (VM). This was representative in terms of mechanical and optical setup, but with only one functional LED channel equipped with a relatively bright 660\,nm LED. Given that all employed materials in NI-CU show a mostly flat wavelength dependency from operational wavelengths down to visual red, the assumption was justified that measurements of the illumination pattern performed at 660\,nm would be representative of the NIR illumination produced at the actual NI-CU wavelengths. The choice of the visual LED and measurements at room temperature allowed the use of a standard charge-coupled device (CCD) detector, making the experimental setup much easier and cost effective. By reducing the measurement distance between NI-CU and the camera by a factor of two some further intensity gain was realised without compromising the accuracy of the measurement. Any sharp reflections or caustics produced by light hitting sharp edges inside NI-CU were expected to be detectable by this setup -- and with a focus on small-scale variations measurements were carried out over the area of the CCD only, corresponding to about a third of the full field of view.

\begin{figure}[h]
\begin{center}
\includegraphics[width=1.0\columnwidth]{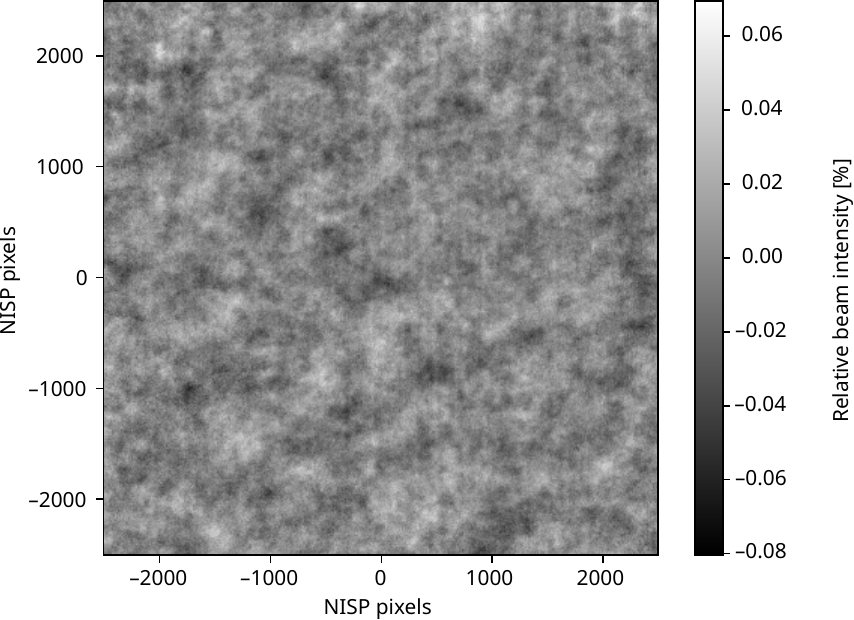}
\end{center}
\caption{Fully calibrated 88\,mm\,$\times$\,88\,mm central section of the NI-CU beam as measured for the VM (with an LED at 660\,nm), corresponding to 4900\,pixel\,$\times$\,4900 NISP pixels, about one third of the NI-CU illuminated field.
\label{fig:NICU_median_image}}
\end{figure}

    Given that the size of the target area was still a multiple of the CCD detector dimensions and that the NI-CU illumination principle involving scattering from a tiny Spectralon target still applied, absolute flux levels were still small, requiring long integration times of many single frames in order to beat noise and arrive at reliable measurements of the true uniformity of the NI-CU illumination pattern. The combined and coadded measurement mosaic of the centre of the NI-CU illuminated area -- corresponding to about one third of the NI-CU field -- after removal of large-scale variations is shown in Fig.~\ref{fig:NICU_median_image}. With this image as a basis a two-dimensional amplitude spectral density of variations was calculated, which showed that the visible structure is at or below a level of 0.07\% peak-to-valley for all scales, after accounting for the details of the measurement process. These variations are likely due to residual micro-roughness of the spectralon surface, for which no specific requirements existed. The maximum of small-scale variations arises at scales of $\sim$\,10 pixel and is smaller on all other scales, including those of the full field-of-view. Hence the NI-CU homogeneity requirements are fulfilled.

During the qualification and verification campaign of the NI-CU models no major discrepancies or non-conformances were encountered. 
Most of the actual problems were related to the quality of some mechanical parts and application of the black PNC coating, but could be resolved without significant delays in the course of NI-CU manufacturing.
Together with the previous space qualification of the infrared LEDs this resulted in the acceptance of all three deliverable models (EQM, FM, FS) and final integration of EQM and FS into the respective NISP models.
No difficulties were encountered during instrument-level tests. Required homogeneity and absolute levels of illumination could be achieved with good agreement between initial estimates for electrical drive parameters and actual values used in the assembled instrument.

\section{NI-CU performance and limits}\label{sec:performance}

The NI-CU flight model was integrated into NISP, and the NISP instrument was tested by itself and as part of the overall \Euclid test campaign before launch. NI-CU already played an important role during these tests, as it was the reference light source inside the integrated NISP instrument to test or verify detector properties and functionality. During \Euclid's in-flight survey operations, NI-CU will be periodically used for pixel response and linearity calibrations \citep{NISP2024}.

\begin{figure*}[ht]
\begin{center}
\includegraphics[angle=0,width=1.8\columnwidth]{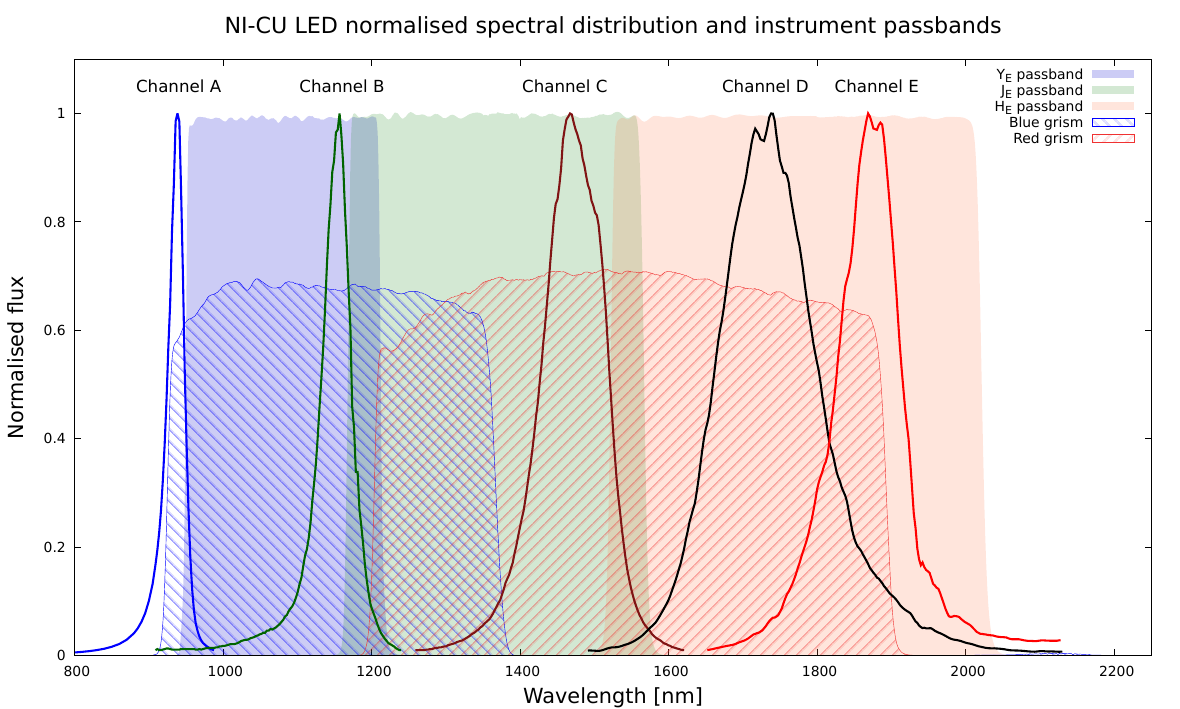}
\end{center}
\caption{Spectra of the five NI-CU channels (lines) at operating temperature (135\,K) in comparison to the NISP instrument filter (shaded areas) and grism passbands (hatched areas). Channel A data has been extrapolated below 900\,nm.\protect\footnotemark}
\label{fig:ledspectra}
\end{figure*}

The FM mechanical and functional properties as determined before launch are summarised in Table~\ref{tbl:FMproperties}. The mechanical parameters are listed mostly as a reference for other projects, while the functional properties are part of the boundary conditions under which NI-CU is operated inside NISP. As part of this, a useful overview of the FM spectral distribution of the five NI-CU illumination channels in comparison to the NISP instrument's filter and grism passbands is shown in Fig.~\ref{fig:ledspectra}.
Each filter passband overlaps with at least one NI-CU channel, and each grism passband with at least two channels. Even though falling slighly outside the \YE-band, channel A can be used together with channel B to interpolate across this passband using ground-based information. The slightly double-peaked and asymmetric structure of the channel D and E spectra is an actual property of the corresponding LEDs.
A more even spacing or an extension towards longer wavelengths might have been preferable, but was in the end limited by the availability of suitable LEDs.

\begin{table}
\caption{NI-CU FM key parameters.} 
\label{tbl:FMproperties}
\centering
\begin{tabular}{l r}
\hline\hline\noalign{\vskip 1pt}

Mass      &  607\,g\\
Length      &  155\,mm\\
Max.\ diameter& 104\,mm\\
Top baffle diameter& 55\,mm\\
Harness lengths&680\,mm\\
Power consumption @135\,K& $\le$\,110\,mW\\
\hline \noalign{\vskip 1pt}
Max drive current $I$& 100\,mA\\
Min drive current $I$, full stability& 10\,mA\\
Min drive current $I$, reduced stability& 1\,mA\\
Min/max PWM duty cycle & 5\%--50\%\\
Drift over 1200\,s, channels A,B,C,E & $\le$\,0.2\%\\
\quad (for drive current 10--100\,mA)&\\
RMS over 1200\,s w/o drift, all channels & $\le$\,0.1\%\\
Stabilisation time & $\lesssim$\,20\,s\\
Spatial homogeneity, RMS & $\lesssim$\,0.15\%\\
\hline
\end{tabular}
\tablefoot{
 The temporal drift is bound by the drift of the driving power supply in NISP's ICU. Its rating is 0.1\% stability at 10\,mA, proportionally decreasing below this.
}
\end{table}

\footnotetext{Measurement of flight LED spectral emission characteristics at cold was limited by the setup to $\geq$\,910\,nm. The characteristic shorter wavelength tail was extrapolated afterwards, using knowledge from earlier measurements of similar LEDs at shorter wavelengths.}

NI-CU's LEDs are mostly operated with drive-currents $I$=10--100\,mA and with a pulse-width-modulation duty-cycle 5--50\%,\footnote{PWM {\em frequency} is nominally 4882\,Hz\,$\pm10\%$. For frequencies this high the systematic noise in pixels -- being sequentially read out -- receiving one more or one fewer pulse over a frame of 1.4\,s is lower than 0.015\%, and hence truly negligible. At lower frequencies already of order 1\,kHz this noise would induce variations of order 0.1\%, a level already similar to temporal stability and spatial homogeneity requirements.} resulting in a dynamic range of 100, a factor ten larger than originally required. Even lower duty cycles are in principle possible, but will stop being rectangular at some point. In combination with potential ageing effects of the power supply this would impact predictability of LED fluence.
The LEDs can, however, be driven with substantially smaller currents down to 1\,mA albeit with an impact on temporal stability (see below).

\begin{figure*}[!ht]
\begin{center}
\includegraphics[angle=0,width=\textwidth]{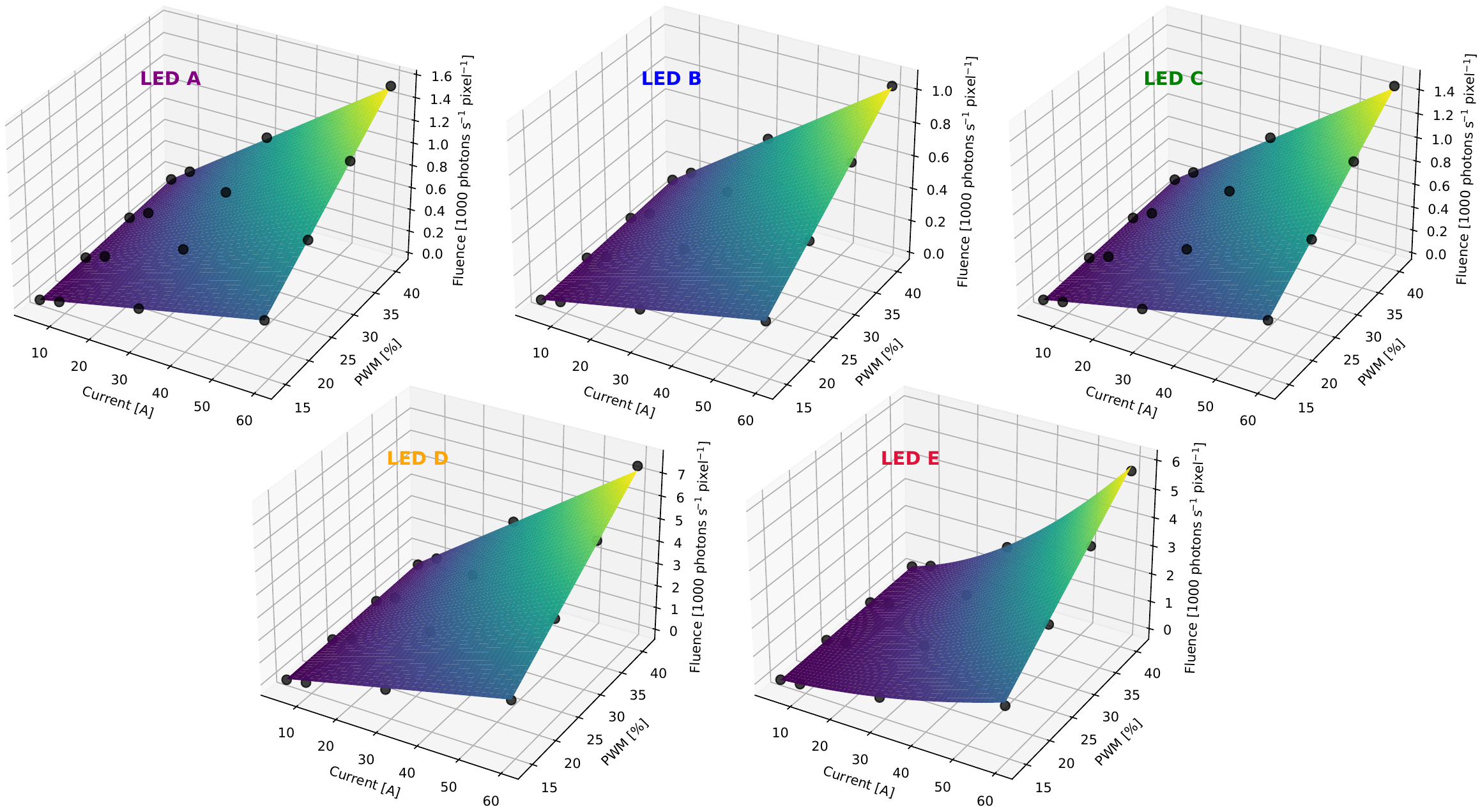}
\end{center}
\caption{
Mean LED fluence at operating temperature (135\,K) expressed in photons per second per detector pixel, as function of driving parameters current and pulse-width-modulation duty cycle (PWM). The points are measurements, the surfaces approximating parameterised functions as listed in Table~\ref{tbl:lut_data}. We note that the flux axes have different scales. 
}
\label{fig:LUT}
\end{figure*}

The measured fluences across the $I$--PWM parameter space vary somewhat between LED types. Measurements -- at the NISP focal plane -- in the core of this parameter space and a functional approximation for easier interpolation are shown in Fig.~\ref{fig:LUT}, with the parameters of the approximating function $f(I,\mathrm{PWM})$ provided in Table~\ref{tbl:lut_data}. 

Except for channel~D all LEDs provide excellent stability in their light output over $\le$1200\,s, as measured for the FM and FS LEDs in a dedicated setup using stable photodiodes during the qualification campaign. The drift is $\le$\,0.2\% over this timespan (for LED~D the drift is $\sim$\,0.6\%\footnote{The primary hypothesis for fluence drift is that variations are primarily due to temperature changes of the semiconductor die. For LED~D however, despite constant ambient temperatures and measured forward voltage, the light output drifted for both FM and FS LED specimen. This is with identical packaging and similar power consumption as the other LED types. We conclude that the small drift is therefore not due to bulk temperature effects, but is likely related to the composition of the LED die, which differs from all other channels. However, this LED will not be used for linearity calibration, only for detector flatfields, so the drift is acceptable.}) and the rms variations after subtraction of a linear drift component are $\le$\,0.1\%. Formally, this is only guaranteed after a warm-up time of 180\,s, but the FM acceptance tests showed stable flux already after typically $\le$\,20\,s. The deposited energy by any LED is small and the thermal coupling to its environment as planned is very good.

All LEDs have been fully tested during development with drive currents down to 1\,mA, and they can also be safely operated at these low currents. The temporal stability however should be expected to decrease. NISP's ICU as the driving power-supply has a rating of 0.1\% stability over 1200\,s, but only down to 10\,mA. Below this current the temporal drift can increase proportionally. While some small-ish level of drift has negligible impact on creating for instance flat-field images for NISP, where the main goal is to inject predictable levels of charges into NISP pixels, larger drifts would start to impact the use of low currents for non-linearity calibration purposes. During NISP linearity calibration 5\,mA are regularly used, and special modes using 1\,mA for very low-level fluences have also been implemented -- but not for linearity calibration purposes.

\begin{table}
\caption{Functional approximation of NI-CU fluence.
} 
\label{tbl:lut_data}
\centering
    \addtolength{\tabcolsep}{-0.1em}
\begin{tabular}{c D{.}{.}{4} D{.}{.}{4} D{.}{.}{4} D{.}{.}{4} D{.}{.}{4}}
\hline\hline\noalign{\vskip 1pt}
Channel & {\it v} & {\it w} & {\it x} & {\it y} & {\it z} \\
\hline\noalign{\vskip 1pt}
A&	-0.6124&	0.3972&	0.6266&	0&	0\\
B&	-0.4845&	0.1185&	0.4277&	0&	0\\
C&	-0.5032&	0.4013&	0.5918&	0&	0\\
D&	-3.2184&	-6.2664&	3.1285&	0&	0\\
E&	-4.6130&	-0.8297&	0.8376&	-0.00055&	0.02771\\
\hline
\end{tabular}
\tablefoot{
Functional approximation of the measured NI-CU fluence, as function of current ($I$ in mA) and pulse-width-modulation duty cycle (PWM in \%) for all five LED channels. The fluence is the prediction per average NISP pixel and per photons per second. These ad-hoc models are of the form $f(v\times I + w\times \mathrm{PWM} + x\times I\times \mathrm{PWM} + y\times I^3 + z\times I^2 \times \mathrm{PWM})$, where $v$, $w$, $x$, $y$, and $z$ are fit parameters across the central part of the $I$ and PWM parameter space. The last two terms are non-zero only for LED~E. The resulting relations are shown in Fig.~\ref{fig:LUT}.
}
\end{table}


\section{Lessons learned, improvements, and discussion}
\label{sec:lessons}
Now after the completion and launch of \Euclid we find it well worth to summarise lessons learned while designing, building and testing NI-CU, for other projects to take into account. (i) The decision to use LEDs as a somewhat novel -- in terms of spaceflight heritage -- light source resulted in technical challenges both during the design and test phase. 
(ii) We would like to address a few points with respect to product and quality assurance as well as interface management. 
(iii) We collected a number of points of possible improvements or enhancements of the NI-CU concept that could be implemented for similar projects with more stringent requirements on absolute calibration or dynamic range.

\subsection{Technical lessons learned}
\label{subsec:techlessons}
While the development of NI-CU proceeded with only few surprises of no real impact on the unit's performance or manufacturability, the following points could help similar projects to a smoother start.

{\bfseries Off-the-shelf vs.\ custom LEDs:}
Selection and qualification of LEDs should start as early as possible. The risk with `commercial off-the-shelf' (COTS) devices is not necessarily that they will not work in space, but qualification will be time consuming and costly, specifically without extensive space legacy. Deciding on and trusting a set of LED devices early during the project helps to limit the subsequent test efforts. Procuring the different raw LED dice and having them assembled into uniform packages by a single service provider streamlined and reduced the qualification scope. Specifically, the uniform build-quality reduced uncertainties from five different LEDs to one. 

{\bfseries Radiation tests, cold vs.\ warm:}
Radiation testing showed significant differences between the individual wavelength channels. This had to be expected because of the range of semiconductor materials involved. Early testing on COTS devices gave a good orientation on performance loss to be expected and helped to make sure that the chosen devices would provide sufficient flux over the mission lifetime. Testing of the final, uniformly packaged devices for the purpose of formal qualification was then done later during the project and did not result in any rejections or surprises. The question if radiation testing should be performed at cryogenic or at room temperature should be discussed and decided early, as the hardware and logistic effort for cryogenic testing will be significantly larger. NI-CU LEDs were tested in warm conditions and there was some remaining risk of underestimating the radiation damage due to non-representative damage annealing taking place between radiation exposure and testing at warm. This risk was seen as acceptable since both the applied radiation doses were factor $\sim$\,2 larger compared to expected conditions and all LED types remained functional after radiation. The substantial margins in drive current range were then available to potentially compensate an underestimated level of damage.

{\bfseries Simple optical layout, complex predictions:}
Regarding the optical layout of the calibration unit, the design based on a diffusor patch and carefully designed apertures turned out to be fully sufficient and very simple in terms of mechanics and optics. The shape of the projected illuminated field could be predicted fairly well based on simulations, while absolute flux levels turned out to have more uncertainty. Here the actual angular illumination pattern and flux levels of the LEDs under operational conditions are not so easy to predict. We recommend to perform early measurements on breadboard level as a starting point. This activity occurred quite late in NI-CU development, leading to some uncertainties on the flux levels to be achieved. The introduction of an additional reflector tube option for the LEDs bore from this uncertainty and could only be removed when first measurements on the engineering and qualification model were available. 
The ideal measurement setup would include a replica of the instrument's focal plane and a large cryogenic chamber -- most likely not readily available. Measurements by proxy should be used cleverly and whenever possible. This means to for example measure the output flux closer to the calibration unit's output than at the original focal plane distance or to use brighter, short wavelength LEDs to assess the dilution factor introduced by the diffusor patch. Cross-calibration between different wavelengths can then take place in a simplified setup. 

{\bfseries Constant need for critical reviews:}
Keeping an open mind to not only see the advantages of LEDs vs. tungsten filaments, but also possible disadvantages, would have helped to deal earlier with the potential issue of light excitation due to susceptibility to electromagnetic interference. When conducting reviews, it helps tremendously to include some reviewers which are {\em not} too familiar with the topic. They will tend to ask the seemingly most harmless questions that turn out to be surprisingly difficult to answer.

{\bfseries Early thermal concept for stability:} 
Last but not least the mechanical and thermal design -- at least in the case of NI-CU -- was developed more or less in parallel to refining the optical concept of the unit. The special circumstances of the glue interface the instrument's SiC required some extra effort on the part of material choice, but did not threaten to be a show-stopper at any time. Starting out with a solid and not overly complex mechanical concept from the beginning and moulding it to its final version worked out well in our case.

\subsection{Organisational \& management lessons learned}
\label{subsec:orglessons}
NISP is a `proper European instrument', as its development and manufacturing was spread over academic and industry partners in at least six countries, plus NASA-JPL, plus ESA as the overseeing customer.
Organisational structures were defined substantially before actual hardware activities for NISP commenced. However, they defined which subsystem developments were later assigned to which consortium partners and where managerial and technical interfaces boundaries were drawn. 

{\bfseries Signal and communication interfaces:}
In the case of NI-CU, the development of the light source was technically decoupled from the development of the required current source and sensing electronics, allocated to the NISP ICU. This resulted in an interface on analogue signal level with -- at the start of hardware development -- only roughly defined requirements regarding absolute values, accuracy, stability, and dynamic range. Most of them could only be narrowed down after initial LED selection and refinement of the illumination concept, that is after fairly certain establishment of the relation between electrical driving parameters and resulting photon flux at the \ac{FPA}.

This led to several instances where electronic requirements had to be re-discussed between the two involved development partners, and then agreed and implemented in the ICU side. At the same time the full development and production of NI-CU ground-support equipment, that is a fully compliant LED driver and housekeeping electronics needed early at least for test purposes in the lab, was tasked to the NI-CU development team. This led to a duplication of efforts.

Avoiding this is not simple. Drawing the interface boundary not in the analogue, but the digital signal domain, would have been interesting, but would require drive electronics by the NI-CU team to be integrated into the flight ICU. Without a detailed control of architecture and components this is challenging to carry out in practice. Similarly, an early provision of drive electronics by the ICU team was not compatible with the NI-CU development timeline.

While the responsibility for the performance critical parameters of the LED driver such as leak current, stability, and dynamic range clearly needs to stay with the team tasked with light source development, a much closer feedback loop for discussing the parameters that are in flux between the two parties seems to be the best option. 
In these communication interfaces, one should not only focus on communication between hardware components but also between the engineers tasked with their development. Even engineers with the same background sometimes interpret seemingly unambiguous requirements in different ways. 

To generalise the experience: choose your interfaces wisely, if possible align them with technical boundaries rather than organisational or political ones, and all interfaces that are technically still in flux require increased attention and a high-frequency cadence of communication between teams. 

{\bfseries Sensible qualification assessment:}
One should critically review the application of standard test and qualification procedures to a hardware development based on `new' components regarding their use in space. In principle, space engineering has developed very valuable test criteria from the experience gained over the past 70 years, following the sensible dictum ``Test as you fly or fly as you test''. 

NI-CU's LEDs had no space heritage, their qualification for a $\sim$\,5-year mission at cold temperatures according to above `product assurance/quality assurance' (PA/QA) principle required to included a 5-year cold storage test -- which obviously is not possible inside a $\sim$\,2-year qualification campaign. Only after intensive discussions this was reduced to a gentle cooldown to operational temperature under vacuum, unbiased storage for a duration of three months, a gentle warmup to room temperature, and subsequent test for any failures or significant performance changes. Before this, however, samples of these LEDs had already been subjected to much harsher treatment, thermal cycling with rapid cooldown and warmup rates, vibrational loads, electrical tests over a whole range of different temperatures -- without failure. Unsurprisingly, the cold storage test showed no deteriorated properties. As predicted, the cold storage test sample have fared better than would have a comparison group just stored at room temperature for the same duration, due to inhibited diffusion processes at semiconductor level. The LEDs in the case of NI-CU had obviously spent a substantial time at room temperature already,\footnote{Until the time of writing this paper, an assumed speed-up of ageing by a factor $\times$2 for every 10\,K temperature increase would translate then $>$\,6 years of storage at room temperature to $>$\,200\,000 (!) years of cold storage at --140\,K.} even higher temperatures would be irrelevant to the actual operation conditions. 

The crucial point is to clarify early on which tests impose {\em in effect} higher or lower loads on a component, and which mechanisms can be invoked from a PA/QA point of view to declare a property successfully qualified. While clearly the above dictum stems from covering previously {\em unknown} effects -- these are still bound by physics.

The much more aggressive thermal and electrical cycling and (mis-)treatment of LEDs during the various other qualification tests was clearly going beyond the test parameter space of the cold storage test, and therefore superior in their significance and ability to instil trust in the selected LEDs. Towards the end of the project, but only in retrospect, there was widespread agreement that the cold storage test was indeed unnecessary. 

As a takeaway regarding product assurance it is paramount to not blindly apply the seemingly closest appropriate standard. 
We strongly recommend to review the rationale behind all test procedures and to make sure each test is indeed useful to address potential failure mechanisms. Pragmatism can help to define a more suitable test strategy -- which could mean to intentionally attempting to break things in order to investigate potential failure modes.\footnote{During initial NI-CU development, LEDs were repeatedly dropped into liquid nitrogen and heated up with hot air, exceeding any thermal shocks during the actual mission by orders of magnitude. They continued to work, immediately giving confidence that normal operations would not come anywhere close to the LEDs thermo-structural limits.} Engaging in a substantial test campaign {\em solely} for formal reasons but without any given physical motivation will not provide progress.

\subsection{Potential improvements}
\label{subsec:improvements}

For a period during the initial development process the requirements were in flux, due to a still settling calibration concept. This initiated discussions of a variety of approaches, and options that later were not implemented. We think two of these central discussions are valuable to describe for future projects.

\subsubsection{Absolute flux calibration}\label{sec:abscal}
During the definition phase of the NISP and NI-CU requirements it was decided that an absolute calibration of detector response by illumination with precisely known flux levels, would not be necessary. Temporal stability of NI-CU flux during individual calibration sequences, approximate spatial uniformity of the illumination pattern for the duration of the mission, and predictable linearity had clear priority. Since the LEDs are always driven with a stable and very precisely controlled current, monitoring the resulting forward voltage provides a diagnostic time-resolved measure for the overall electrical power consumption, and hence light output. As this can be done to very high precision, and in the case of NI-CU the parameter is constantly read and downlinked as part of the housekeeping data, the forward voltage is a very sensitive proxy-parameter for light emission -- sufficient for the required temporal monitoring of LED fluence.

To achieve {\em absolute} monitoring and calibration of the focal plane sensitivity would require either a very stable calibration source or reliable in-orbit measurement of the calibration source's output. This measurement itself would then again have to be accurate and stable during the mission lifetime, resistant to ageing effects or radiation damage. Such a calibration strategy would allow the users to assess the degradation of the focal plane assembly and of components in the optical path and its mirrors, lenses, and filters, when combined with observations of stable reference stars.

A technical solution for this challenge presents a classical chicken-egg-problem. Detectors which could be used to monitor the NI-CU flux normally suffer from the same damage and ageing mechanisms as the LEDs used in NI-CU. These detectors would typically be photodiodes, similarly being susceptible to radiation damage. Other types of detectors such as lead sulfide photoresistors, thermopiles, or radiometric solutions are either prone to sensitivity drifts as well or are not suitable for the considered flux levels.
Therefore any observed change of focal plane properties could be attributed either to the focal plane itself, a change of the calibration source output, a change of the calibration source monitoring detector sensitivity, or any combination of those.

A resort from this dilemma would likely hinge on protecting the monitoring detector from radiation damage or being able to predict its long-term performance with high accuracy. The latter could be based on a good knowledge of the dependency of the detector sensitivity on absorbed radiation and a separate radiation monitoring device. Moreover, it would also require a non-ageing power source, which is also non-trivial to achieve. The accuracy of such a solution would probably be severely limited by the non-linearity and energy-dependency of the sustained radiation damage of all components involved.

If a relationship between absorbed radiation dosage and LED output flux could be well established in the laboratory, multiple sets of LEDs per wavelength channel with different amounts of shielding applied might also permit to computationally cancel out radiation effects to some degree.

A technically more complex but probably most reliable solution would be to calibrate the monitoring detector of NI-CU with a separate, more radiation-robust light source from time to time. This could be a light source with predictable ageing effects and less radiation damage susceptibility, for example a tungsten filament. For such a source, degradation of the output flux mostly depends on operating time. If such a source was only used for occasional calibration purposes of the monitoring detector, good long-term stability might be feasible. An even more advanced variation of this approach would solely use an external, well-known extended celestial light source -- in the case of NISP only very wide angle sources, that is only the Sun and Moon would in principle make sense to calibrate the large field-of-view\footnote{Mars, Jupiter, and Saturn have angular diameters of only $\sim$\,5\arcsec--50\arcsec or 17--170 NISP pixels across, while the Moon and Sun would cover $\sim$\,40\% of the FPA area.} -- but this would add substantial complexity on hardware level. For example, \Euclid has to always point away from Sun, Earth, and Moon for stability purposes. It would have required a very different telescope and instrument design to make Sun or Moon usable for detector calbration. \Euclid already uses stars from its survey in a self-calibration approach to model the large-scale sensitivity variations of the FPA. This approach however is not well suited for pixel-to-pixel variation calibration.

Any requirement for absolute calibration on long time scales therefore results at least in additional components, mass, electronics, and algorithms development. It should be well justified in the sense of being mission critical.

\subsubsection{Increase of dynamic range}\label{sec:dynamicrange}

Depending on the required accuracy and reproducibility over the mission duration, the dynamic range for a single set of LEDs is limited to $\sim$\,1:200. This results from the combination of the current source's dynamic range ($\sim$\,1:10--1:20 in the case of $<$0.1\% stable operations of the NISP ICU) and the maximum and minimum PWM duty cycle, respectively ($\sim$\,1:10). An improved current source design or less demanding requirements on linearity and long-term reproducibility might result in a somewhat larger range, though. This would require design effort on the electronics side of the system without touching the NI-CU hardware itself.

A further increase of the dynamic range at the cost of volume and component count could be achieved by adding additional sets of LEDs to the calibration unit. Instead of having just one nominal and possibly one redundant LED per wavelength, multiple LEDs per channel with different illumination properties would be employed. The term `illumination property' here refers to the conversion of electrical power to flux level at the detector. An identical second, or even third, LED per channel could be equipped with a sort of attenuator or illuminate the diffusor patch from a larger distance or from a less favourable angle. As a result, the same ICU drive signal could produce different illumination levels for the different LED sets.

In the case of NI-CU, a second set of LEDs could either have been squeezed into the LED holder as designed if a slightly larger volume could be accommodated or built into a separate holder, stacked on top of the first one. A more sophisticated solution could consist of daisy-chaining Ulbricht style integration spheres, as planned for the Roman WFI calibration source (Joshua E.\ Schlieder, pers. comm.). For \Euclid, however, this could not have been accommodated into the available volume of NISP.

Independent of the chosen solution, additional light sources will require additional electrical connections and more effort on the part of the control electronics. Both should be taken into account early during the design phase. As with a requirement for absolute calibration, the need for an extended and reproducible dynamic range should be well justified.

\section{Conclusion}\label{sec:final}
In summary, NI-CU's development was in principle straightforward, not only in retrospect. Given the central requirements of time-stable illumination at several well-defined wavelengths, but variable fluence, LEDs are clearly the best choice as lightsources. Their very low energy dissipation -- and therefore stable thermal conditions -- and fluence-independent emitted spectrum would have been very challenging if not impossible to reach with tungsten filament lamps. The ability to monitor LED fluence levels with voltage sense lines is a unique property, that enables a continuous, fast, and simple approach to provide monitoring or housekeeping data -- both during development and now in flight.

In hindsight, even though test-LEDs basically already clarified very early on that LEDs would function well in the cryo-vacuum and radiation environment at L2, the full LED qualification process was substantially more extensive (and expensive) than initially envisioned. We hope that some of these steps will not be necessary in the future, now that after NI-CU there is substantially more `legacy' for NIR-LEDs in space. And that at some point limited-qualification COTS LEDs might become acceptable in similar projects, even if not all manufacturing processes are fully known. Since in principle LEDs are inexpensive, the difference in cost to custom-manufacturing and a full qualification can well be a factor (!) of 20--50.

NI-CU development was ultimately very successful, leading to a very capable calibration light-source. The FM integrated into NISP was already used during instrument ground tests to check out and provide reference data of the NISP FPA -- and is now orbiting at L2, supporting NISP in \Euclid's 6-year survey of the Universe.

%
%

\begin{acknowledgements}

The authors would like to thank Coryn Bailer-Jones (on Gaia calibration), Jörg-Uwe Pott and Robert Harris (on MICADO calibration approaches), and Joshua E.\ Schlieder (on the Roman WFI calibration source) for interesting information and discussions. We want to further thank Thomas Blümchen for his work on an early version of NI-CU, Armin Böhm and the mechanics workshop and Ulrich Grözinger at MPIA for their support, the HIT facility at Heidelberg University Hospital and specifically Martin H\"artig, the team at {\em von Hoerner \& Sulger} for their work, and Frank Sackenheim, Stefan Binnewies, and Josef Pöpsel from Capella Observatory for loaning us their STX-16803 camera.

We specifically acknowledge funding of NI-CU development and MPIA's \Euclid-involvement through the Deutsches Zentrum für Luft- und Raumfahrt (German Aerospace Center, DLR) under grants 50\,QE\,1202, 50\,QE\,2003, and 50\,QE\,2303.\\

\AckEC
\end{acknowledgements}

\bibliography{Euclid,paper}

\end{document}